\documentclass[conference]{IEEEtran}
\IEEEoverridecommandlockouts
\usepackage{amssymb}
\usepackage{cite}
\usepackage{graphicx}
\usepackage{array}
\usepackage{multirow}
\usepackage{amsmath}
\usepackage{stfloats}
\usepackage{graphicx}
\usepackage{subfigure}
\usepackage{tabularx}
\usepackage{epsfig,epsf,balance}
\usepackage[caption=false,font=normalsize,labelfont=sf,textfont=sf]{subfig}
\usepackage{setspace}
\usepackage{bm}
\usepackage{textcomp}
\usepackage{algorithmic}
\usepackage{algorithm}
\usepackage{subfigure}
\usepackage{caption}
\usepackage{graphicx}
\usepackage{setspace}
\usepackage{url}
\usepackage{verbatim}
\usepackage{graphicx}
\usepackage{cite}
\usepackage{amssymb}
\usepackage{color}
\usepackage{xpatch}
\definecolor{shadecolor}{RGB}{0,0,255}
\definecolor{blue}{RGB}{0,0,255}

\newtheorem{theorem}{Theorem}

\newtheorem{lemma}{Lemma}
\newtheorem{corollary}{Corollary}
\newtheorem{remark}{Remark}

\makeatletter

\ExplSyntaxOn
\cs_new:Npn \bibColoredItems #1#2
{
	\clist_map_inline:nn {#2} { \cs_new:cpn {bib@colored@##1} {#1} } 
}
\ExplSyntaxOff

\newcommand\bib@setcolor[1]{%
	\ifcsname bib@colored@#1\endcsname
	\expanded{\noexpand\color{\csname bib@colored@#1\endcsname}}%
	\else
	\normalcolor
	\fi
}

\xpatchcmd\@bibitem
{\item}
{\bib@setcolor{#1}\item}
{}{\fail}

\xpatchcmd\@lbibitem
{\item}
{\bib@setcolor{#2}\item}
{}{\fail}
\makeatother

\begin{document}
	
\title{Dynamic Resource Allocation in Distributed MIMO-LEO Satellite Networks
}
 
\author{Qihao Peng, Qu Luo, ~\IEEEmembership{Member,~IEEE}, Yi Ma, ~\IEEEmembership{Senior Member,~IEEE}, 
	Chuan Heng Foh, ~\IEEEmembership{Senior Member,~IEEE}, \\ Pei Xiao, ~\IEEEmembership{Senior Member,~IEEE}, Maged Elkashlan, ~\IEEEmembership{Senior Member,~IEEE} \\
    Rahim Tafazolli, ~\IEEEmembership{Fellow,~IEEE}, George K. Karagiannidis,  ~\IEEEmembership{Fellow,~IEEE}.

		\thanks{Q. Peng, Q, Luo, Yi Ma, C. Foh, P. Xiao, and R. Tafazolli are affiliated with 5G and 6G Innovation Centre, Institute for Communication Systems (ICS) of the University of Surrey, Guildford, GU2 7XH, UK. (e-mail: \{q.peng, q.u.luo, y.ma, c.foh, p.xiao, r.tafazolli\}@surrey.ac.uk). (\emph{Corrosponding Author: Qu Luo})} \\
        \thanks{M. Elkashlan is with the School of Electrical Engineering and Computer Science of Queen Mary University of London. (e-mail: m.elkashlan@qmul.ac.uk).} \\
        \thanks{George K. Karagiannidis is with the Department of Electrical and Computer Engineering, Aristotle University of Thessaloniki, Greece. (email: geokarag@auth.gr). }}

	
	\maketitle
	
\begin{abstract}
	This paper characterizes the impacts of channel estimation errors and Rician factors on achievable data rate and investigates the user scheduling strategy, combining scheme, power control, and dynamic bandwidth allocation to maximize the sum data rate in the distributed multiple-input-multiple-output (MIMO)-enabled low earth orbit (LEO) satellite networks. However, due to the resource-assignment problem, it is challenging to find the optimal solution for maximizing the sum data rate. To transform this problem into a more tractable form, we first quantify the channel estimation errors based on the minimum mean square error (MMSE) estimator and rigorously derive a closed-form lower bound of the achievable data rate, offering an explicit formulation for resource allocation. Then, to solve the NP-hard problem, we decompose it into three sub-problems, namely, user scheduling strategy, joint combination and power control, and dynamic bandwidth allocation, by using alternative optimization (AO). Specifically, the user scheduling is formulated as a graph coloring problem by iteratively updating an undirected graph based on user requirements, which is then solved using the DSatur algorithm. For the combining weights and power control, the successive convex approximation (SCA) and geometrical programming (GP) are adopted to obtain the sub-optimal solution with lower complexity. Finally, the optimal bandwidth allocation can be achieved by solving the concave problem.
     Numerical results validate the analytical tightness of the derived bound, especially for large  Rician factors, and demonstrate significant performance gains over other benchmarks.
\end{abstract}	
	
\begin{IEEEkeywords}
		Distributed MIMO-LEO networks, cooperative reception, user scheduling, dynamic bandwidth.
\end{IEEEkeywords}

\section{Introduction}
Satellite communication has emerged as a pivotal technology to achieve global seamless connectivity, particularly in scenarios where terrestrial networks are infeasible, such as remote areas, maritime zones, and disaster recovery operations \cite{you2020leo,wu2024large}. The heterogeneous integration of satellite-terrestrial networks enables a ubiquitous coverage paradigm that extends the core service dimensions of 5G systems into traditionally underserved geographical domains. Against this background, satellite communications have drawn extensive attentions from the industry, such as SpaceX \cite{SpaceX} and OneWeb \cite{OneWeb}. Meanwhile, the transparent payload-based satellite links were standardized in 3GPP Release 17 \cite{3GPP}, aligning with the vision of integrated space-terrestrial ecosystems in 6G.

Benefiting from the multiple-input-multiple-output (MIMO) \cite{2014MIMO,peng2022resource}, satellites equipped with multi-antenna systems can simultaneously serve multiple users via shared frequency and bandwidth resources, and thus extensive contributions have been devoted to the investigation of MIMO-enabled satellite networks \cite{li2021downlink, you2020massive, KexinChanel, shen2022random,gao2021sum}. Specifically, the authors in \cite{li2021downlink} investigated downlink MIMO beamforming based on statistical channel state information (CSI), aiming to address the challenge of acquiring instantaneous CSI due to the long-distance propagation inherent in satellite communications.  Then, Li \emph{et al.} devised the transmission design to serve a large number of users and proposed a user group strategy relying on space angle \cite{you2020massive}. To address the challenges of severe path loss and imperfect CSI, a two-stage channel estimation is proposed in \cite{KexinChanel}, which achieves near-optimal performance to the maximum mean square error (MMSE) with lower complexity. To ensure robust communication over high-mobility satellite channels, an orthogonal time-frequency space modulation-enabled satellite system was investigated in \cite{shen2022random}. Furthermore, the benefit for combining non-orthogonal multiple access with low earth orbit (LEO) satellite communication systems was explored \cite{gao2021sum}. It is worth noting that the existing studies in \cite{li2021downlink,you2020massive,KexinChanel,shen2022random,gao2021sum} primarily focused on the investigation of single-satellite systems.

Given that multiple satellites may be visible to each user, it is worth exploring the cooperative gain among multiple satellites. Therefore, distributed MIMO-enabled satellite networks have drawn extensive attentions \cite{OptMIMO, richter2020downlink, li2021analysis,li2021capacity,abdelsadek2023broadband,roper2022distributed,xiang2024DLmassive,xiang2024ULmassive,abdelsadek2022distributed}. In particular, to investigate the cooperation among multiple satellites, the optimal MIMO capacity can be achieved by devising the locations and antenna elements of satellites and users \cite{OptMIMO}. Richter \emph{et al.}  exploited the cooperative transmission of two satellites and revealed that the cooperative gain can be obtained only when two users are far away from each other \cite{richter2020downlink}. Subsequently, the channel capacity of distributed satellites equipped with uniform linear arrays (ULAs) was rigorously analyzed in \cite{li2021analysis}, and this framework was extended to satellites with uniform circular arrays (UCAs) in \cite{li2021capacity}. Similarly, the closed-form expressions for spectral efficiency were derived for distributed massive MIMO  systems \cite{abdelsadek2023broadband}, providing the theoretical foundation for distributed satellite systems. For signal processing, the distributed precoding and equalization were proposed in \cite{roper2022distributed}. To solve the asynchronous issues in time and frequency, the authors of \cite{xiang2024DLmassive} devised spatial linear receive processing for signal extraction with time and frequency compensations, and further designed a Riemannian conjugate gradient-based detection algorithm \cite{xiang2024ULmassive}. To provide seamless connectivity for users, the power allocation and handover management based on artificial intellengece were investigated in \cite{abdelsadek2022distributed}. Based on the existing work, it is been shown that distributed MIMO-enabled satellite networks will be a promising solution for 6G.

However, compared to terrestrial networks, distributed MIMO-LEO networks remain a critical research frontier constrained by the fundamental limitations, such as imperfect CSI and spatially correlated channels. Specifically, one of the critical challenges is that the inherently severe path loss, compared to the noise background, results in inaccurate CSI \cite{KexinChanel,lin2023low,chang2022csi}. This necessitates a rigorous characterization of channel estimation error impacts on achievable data rates. Furthermore, owing to long-distance propagation, it is challenging for LEO satellites to distinguish the densely distributed users, which induces severe inter-user interference and impedes interference management. 

To bridge the above-mentioned gaps, our work makes three pivotal contributions. Firstly, we investigate the cooperative reception of satellites by deriving the closed-form expression during the uplink transmission, providing comprehensive performance analysis for distributed MIMO-LEO satellite networks. Secondly, based on the given analysis, the achievable data rate is fundamentally limited by inter-user interference, necessitating user scheduling and resource allocation to guarantee the users' quality of service. Finally, to maximize the sum data rate, we jointly optimize the user scheduling strategy, the combination strategy, transmission power, and bandwidth allocation to fully unleash the cooperative gain. Our main contributions are summarized in the following:
\begin{enumerate}
	\item  Considering a typical Rician fading channel comprising both line-of-sight (LoS) and non-line-of-sight (NLoS) components, channel estimation errors are quantified using the MMSE method, followed by the evaluation of the normalized mean square error (NMSE). It is observed that the estimation error decreases as the Rician factor increases. However, a higher Rician factor may also result in increased NMSE, suggesting that precise CSI acquisition becomes more challenging under strong LoS conditions due to the dominance of noise and reduced multipath diversity.
    \item Based on the estimated channel, the closed-form expression of the lower bound is derived while considering pilot contamination and imperfect CSI, which offers an explicit form for later resource assignment. Through rigorous analysis, it is shown that each user's data rate becomes limited when the Rician factor exceeds 13 dB. Furthermore, the user's interference will significantly deteriorate the system performance owing to the correlated channels, which motivates us to explore the user scheduling scheme.
    \item Based on the given closed-form expression of achievable data rate, we aim to maximize the sum data rate by jointly optimizing the user scheduling scheme, combining weights, transmission power, and bandwidth allocation. To solve this NP-hard problem, we decompose it into three sub-problems, including user scheduling, joint combination and power control, and dynamic bandwidth allocation. Firstly, to reduce the computational complexity of exhaustive search, a low-complexity algorithm based on the DSatur algorithm is proposed. Then, by adopting the successive convex approximation (SCA) and geometrical programming (GP), the combining weights and transmission power can be found. Finally, the dynamic bandwidth allocation can be obtained by solving the concave problem.
    \item Simulation results validate the accuracy of our derivations and analysis. For the user scheduling strategy, it is beneficial to consider a frequency division system in distributed satellite networks, especially for a large number of users. Furthermore, our proposed algorithm is superior to the other benchmarks, verifying the effectiveness of our method.    
\end{enumerate}

The remaining sections are organized as follows. The system model and a comprehensive analysis of channel estimation are presented in Section II. The closed-form expression of the lower bound is derived, and the performance analysis is given in Section III. Then, based on the given lower bound, the user scheduling strategy, combination weights, transmission power, and bandwidth are jointly optimized in Section IV. In Section V, extensive numerical results are presented. Finally, our conclusions are drawn in Section VI.

\section{Satellite System}

\begin{figure}
	\centering
	\includegraphics[width=3.2in]{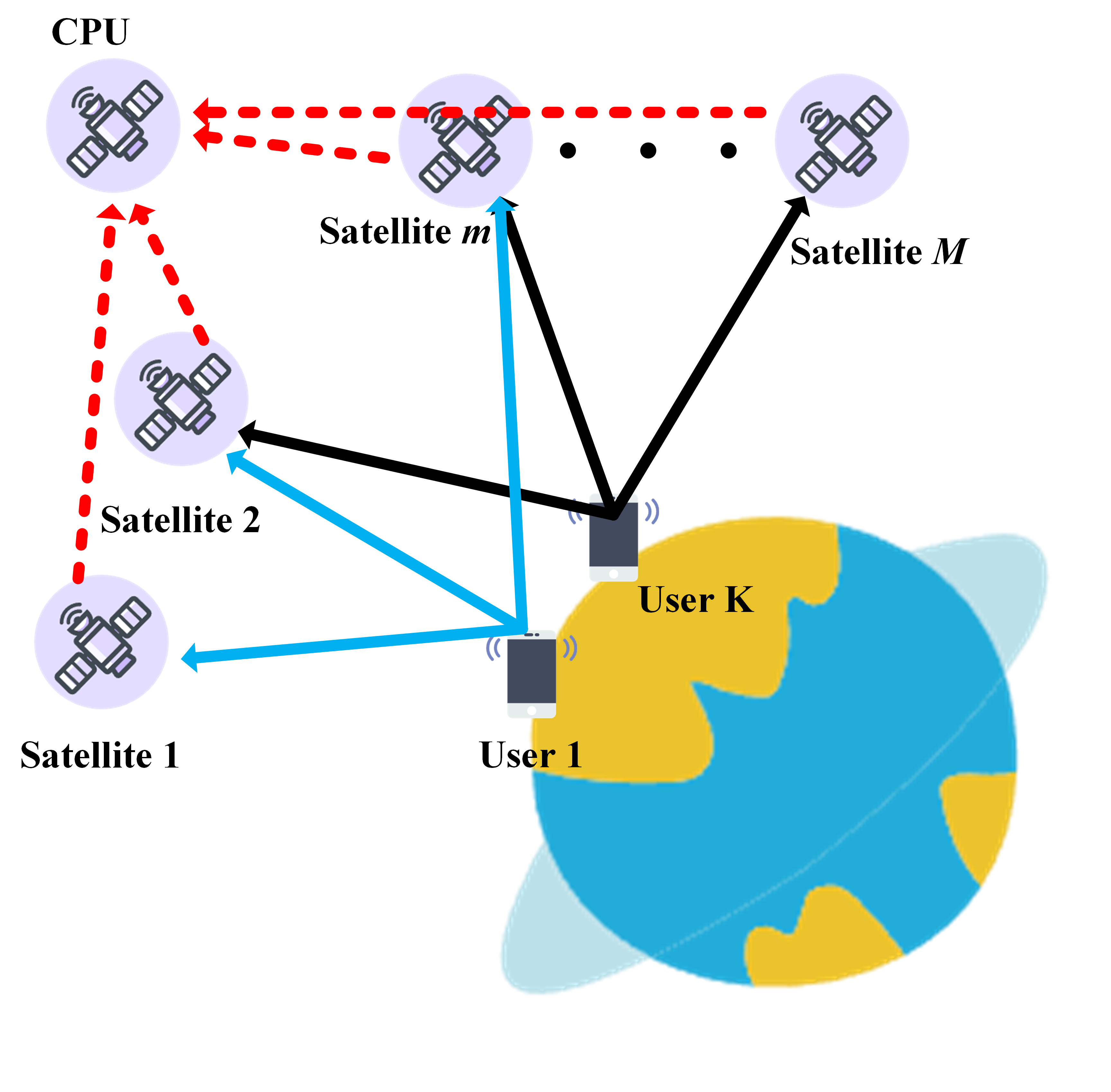}
	\caption{System model for distributed MIMO-enabled satellite networks.}
	\label{fig:system}
\end{figure}
\subsection{System Model}
As shown in Fig. \ref{fig:system}, we consider a distributed MIMO-enabled satellite networks, where \(M\) \(N\)-antenna satellites serve \(K\) single-antenna users. Without loss of generality, we respectively define \(\mathcal{M}\) and \(\mathcal{K}\) as the set of satellites and users, with \(\forall m \in \mathcal{M}\) and \(\forall k \in \mathcal{K}\). Based on the user-centric approach, the set of satellites that provide service for the \(k\)-th user is denoted as \(\mathcal{M}_k\). Then, we suppose that a satellite network operating in frequency division duplexing (FDD) mode with an aggregate system bandwidth of \(B\). Furthermore, to simultaneously deliver high-quality services to multiple users, the central processing unit (CPU) employs a dynamic resource management strategy that jointly optimizes bandwidth partitioning and user association. Specifically, the total bandwidth is partitioned into \(I\) \((I < K)\) sub-bands satisfying \(\sum\nolimits_{i=1}^I B_i = B\), where each sub-band \(B_i\) is allocated to an unique set of users \(\mathcal{K}_i\), i.e., \(\mathcal{K}_i \cap \mathcal{K}_j = \emptyset\) when \(i \neq j\). 

Due to the fixed orbit and known velocity of satellites, we assume that the Doppler shifts can be compensated before transmission \cite{li2021capacity,deng2022holographic,riera2022scalable}. Therefore, the channel \(\mathbf{h}_{m,k} \in \mathbb{C}^{N \times 1}\) from the \(m\)-th satellite to the \(k\)-th user is modeled as
\begin{equation}
    \label{channel}
    \small
    \begin{split}
        \mathbf{h}_{m,k} &=  \sqrt{\frac{{\beta}_{m,k}}{{\bar K}_{m,k}+1}}\Big(\sqrt{\bar K_{m,k}}\mathbf{\bar{h} }_{m,k} + \boldsymbol{\Delta}^{\frac{1}{2}}_{m,k}\mathbf{\widetilde h}_{m,k}\Big),
    \end{split}
\end{equation}
where \({\bar K}_{m,k}\) is the Rician factor related to elevation angle according to the 3GPP standard \cite{3GPP}, \(\beta_{m,k}\) is the large-scale fading factor, \(\boldsymbol{\Delta}_{m,k} \in \mathbb{C}^{N \times N}\) is the correlation matrix, \(\mathbf{\bar{h} }_{m,k}\) and \(\mathbf{\widetilde h}_{m,k}\) are the LoS and nLoS channels, respectively. For the LoS channels, \(\mathbf{\bar{h} }_{m,k}\) based on the uniform planar array (UPA) is given by \cite{shafin2017angle}
\begin{equation}
    \label{LoSchanel}
    \begin{split}
    \small
           \mathbf{\bar{h} }_{m,k} &=  {\mathbf{a}}^x_{m,k}(\phi_{m,k},\theta_{m,k})  \otimes  {\mathbf{a}}^y_{m,k}(\phi_{m,k},\theta_{m,k}), \\
    \end{split}
\end{equation}
where \(\otimes\) is the Kronecker product, \({\mathbf{a}}^x_{m,k}(\phi_{m,k},\theta_{m,k})\) and \({\mathbf{a}}^y_{m,k}(\phi_{m,k},\theta_{m,k})\) are the steering vectors, respectively. The \(n\)-th (\(n \in \{0,\cdots, N_x -1\}\)) element of \({\mathbf{a}}^x_{m,k}(\phi_{m,k},\theta_{m,k})\) can be expressed as \(e^{-j2\pi \frac{d_A}{\lambda} (n-1)\sin(\phi_{m,k})\cos{\theta_{m,k}}}\) and the \(n\)-th (\(n \in \{0,\cdots, N_y -1\}\)) element of \({\mathbf{a}}^y_{m,k}(\phi_{m,k},\theta_{m,k})\)  is \(e^{-j2\pi \frac{d_A}{\lambda}(n-1) \cos(\phi_{m,k})}\), where \(\lambda\) is the wavelength of the carrier frequency, \(N_x\) (\(N_y\)) is the number of antennas of the x-axis (y-axis) direction with \(N = N_x N_y\), \(d_A\) is the antenna spacing, \(\phi_{m,k}\) and \(\theta_{m,k}\) denote the elevation and azimuth angles at which the LoS signal is transmitted from satellite \(m\) to user \(k\). For the NLoS channel, each element of \(\mathbf{\widetilde h}_{m,k}\) follows the independent and identically distributed complex Gaussian random variables with zero mean and unit variance.

\subsection{Channel Estimation}
In a dense network, the number of access users is generally larger than the number of available orthogonal pilot sequences (i.e., \(\tau \le K\)), leading to pilot contamination. Therefore, we denote the set of users that share the same pilot sequence with the \(k\)-th user as \(\mathcal{P}_k \in \mathcal{K}\). If the user \(k\) share the same common pilot with the \(k'\)-th user, then we have \(\mathcal{P}_k = \mathcal{P}_{k'}\).

In the training phase, the received pilot signal at the \(m\)-th satellite is denoted as
\begin{equation}
\label{received_pilot}
\small
\setlength\abovedisplayskip{5pt}
\setlength\belowdisplayskip{5pt}
{{\bf{Y}}^p_{m}} = \sum\limits_{k \in \mathcal{K}} {{{\bf{h}}_{m,k}}\sqrt { p_k^p} {\bf{ q }}_{k}^H}  + {{\bf{N}}^p_{m}},
\end{equation}
where \(p_k^p\) is the pilot power of the \(k\)-th user, and \({{\bf{N}}_{m}^p} \in  {{\mathbb{C}}^{N \times \tau}}\) is the additive Gaussian noise matrix at the \(m\)-th satellite, each element of which is independent and follows the Gaussian distribution of \(\mathcal{CN} \left( {{{0}},{{\sigma^2}}} \right)\) \footnote{Here, we assume that the channel estimation is based on all the bandwidth of \(B\) to acquire the channel state information, thereby facilitating the sub-band allocation.}. By multiplying (\ref{received_pilot}) with \({\bf{ q }}_{k}\), we have
\begin{equation}
\label{received_channel}
\small
\setlength\abovedisplayskip{5pt}
\setlength\belowdisplayskip{5pt}
{{\bf{\hat y}}_{m,k}^p} = \frac{1}{{\sqrt {\tau } }}{{\bf{Y}}^p_{m}}{{\bf{q}}_{k}} = \sum_{j \in \mathcal{P}_k} \sqrt{\tau p^p_{j}}{{\bf{h}}_{m,j}} + {\bf{n}}_{m,k}^p,
\end{equation}
where \({\bf{n}}_{m,k}^p = \frac{1}{{\sqrt {\tau} }}{{\bf{N}}^p_{m}}{{\bf{q}}_{k}}\). Based on (\ref{received_channel}), the MMSE estimate for ${\bf{h}}_{m,k}$ is
\begin{equation}
\label{estimate_gmk}
\begin{split}
\small
{{{\bf{\hat h}}}_{m,k}} 
& = \sqrt{\frac{{\bar K}_{m,k}}{{\bar K}_{m,k}+1}}\sqrt{\beta_{m,k}}\mathbf{\bar{h} }_{m,k} \\
&+ \sqrt{\tau p_k^p}\mathbf{R}_{m,k}\boldsymbol{\Psi}_{m,k} \Big(\sum_{j \in \mathcal{P}_k}\sqrt{\frac{\tau p^p_j\beta_{m,j}  }{{\bar K}_{m,j}+1}}\boldsymbol{\Delta}_{m,j}^{\frac{1}{2}}\mathbf{\widetilde h}_{m,j} \\
&+{\bf{n}}_{m,k}^p \Big),
\end{split}
\end{equation}
where \(\mathbf{R}_{m,k}\) and \(\boldsymbol{\Psi}_{m,k}\) are given by
\begin{equation}
    \label{Rmk}
    \small
    \begin{split}
         \mathbf{R}_{m,k} &= \text{Cov}\{\mathbf{h}_{m,k},\mathbf{h}_{m,k}\} = \frac{\beta_{m,k}}{{\bar K}_{m,k}+1}\boldsymbol{\Delta}_{m,k}\\
         & \triangleq a_{m,k}\boldsymbol{\Delta}_{m,k},
    \end{split}
\end{equation}
and 
\begin{equation}
    \label{Psimk}
    \begin{split}
    \small
        \boldsymbol{\Psi}_{m,k} &= \Big(\sum\limits_{j \in \mathcal{P}_k}{\tau p^p_j \mathbf{R}_{m,j} +\sigma^2\mathbf{I}_N}\Big)^{-1}.
    \end{split}
\end{equation}
Furthermore, the distribution of the estimated channel can be written as
\begin{equation}
    \label{Distributionchanel}
    \begin{split}
    \small
        \mathbf{\hat h}_{m,k} \sim \mathcal{CN}\Big(\mathbf{\bar h}_{m,k},\tau p^p_k \mathbf{R}_{m,k}\boldsymbol{\Psi}_{m,k} \mathbf{R}_{m,k}\Big).
    \end{split}
\end{equation}
The estimation error of \(\mathbf{h}_{m,k}\), denoted as \(\mathbf{e}_{m,k}\), is given by 
\begin{equation}
    \label{Distributionerror}
    \begin{split}
    \small
       \mathbf{e}_{m,k} &= \mathbf{ h}_{m,k}- \mathbf{\hat h}_{m,k} \\
       &\sim \mathcal{CN}(\mathbf{0},\mathbf{R}_{m,k}-\tau p^p_k \mathbf{R}_{m,k}\boldsymbol{\Psi}_{m,k} \mathbf{R}_{m,k}).
    \end{split}
\end{equation} 
The NMSE of the estimated channel can be given by
\begin{equation}
    \label{NMSE}
    \begin{split}
    \small
        \text{NMSE}_{m,k} = \frac{\text{tr}\{\mathbf{R}_{m,k} - \tau p^p_k \mathbf{R}_{m,k}\boldsymbol{\Psi}_{m,k} \mathbf{R}_{m,k}\}}{\text{tr}\{\mathbf{R}_{m,k}\}},
    \end{split} 
\end{equation}
where \(\text{tr}\{x\}\) denotes the trace of \(x\).

\begin{remark}
Considering the LoS channel, i.e., \(\bar{K}_{m,k} \rightarrow \infty\), the estimated channels only have the LoS components, and thus the estimated channel errors become zero owing to the deterministic and known LoS channel. However, in this case, the NMSE of the estimated channel will approach to 1, owing to the significant interference of noise. Furthermore, in the case of pilot length \(\tau \rightarrow \infty\) and no pilot sharing, i.e., \(|\mathcal{P}_k| = 1\), we have \(\mathbf{\hat h}_{m,k} \rightarrow \mathbf{h}_{m,k}\) and \(\mathbf{e}_{m,k} \rightarrow \mathbf{0}\), which implies the estimated channels are perfect. 
\end{remark}

\subsection{Uplink Transmission}
Based on the user-centric approach, the CPU can combine signals from the satellites of the set \(\mathcal{M}_k\) to decode the \(k\)-th user's information. Furthermore, as the \(k\)-th user transmits its signal via the \(i\)-th sub-band, the interference only comes from the users of set \(\mathcal{K}_i\). Then, the CPU combines the signals from satellites in the set \({\mathcal{M}}_k\) to detect the \(k\)-th user's information, which is given by
\begin{align}
\setlength\abovedisplayskip{5pt}
\setlength\belowdisplayskip{5pt}
\small
y_k^d  &= \mathbb{E} \underbrace {\left\{ {\sum\limits_{m \in {\mathcal{M}}_k} {w_{m,k}}{{{\left( {{\bf{ \hat h}}_{m,k}} \right)}^H}{\bf{h}}_{m,k}\sqrt {{p_k^d}} } } \right\}}_{{\rm{DS}}_k}{s_k}  \notag \\ 
&\!+\!\underbrace { \left\{\sqrt {{p_k^d}}{\sum\limits_{m \in {\mathcal{M}}_k} {w_{m,k}}{{{\left( {{\bf{ \hat h}}_{m,k}} \right)}^H}{\bf{h}}_{m,k}}  - {\rm{DS} }_k  } \!\right\}}_{{\rm {LS}}_k}\!{s_k}, \label{kth_statistical_signal} \\
 & + \sum\limits_{k' \in \{\mathcal{K}_i \backslash k\}} { \underbrace {\sum\limits_{m \in {\mathcal{M}}_k}\!\!{w_{m,k}}{{{\left( {{\bf{ \hat h}}_{m,k}} \right)}^H}{\bf{h}}_{m,k'}\sqrt {{p_{k'}^d}} } }_{{\rm {UI}}_{k,k'}}\!{s_{k'}}} \notag \\
 &+ \underbrace {\sum\limits_{m \in {\mathcal{M}}_k}\!\! {w_{m,k}}{{{\left( {{\bf{ \hat h}}_{m,k}} \right)}^H}{\bf{n}}_m} }_{{{\rm {N}}_k}} \notag,
\end{align}
where \(w_{m,k}\) is the weight for the \(m\)-th satellite to the \(k\)-th user with \(\sum\limits_{m \in \mathcal{M}_k}w^2_{m,k} = 1\), \(p^d_k\) is the $k$-th device's power, $s_k$ is the transmitted information, ${\bf{n}}_m$ is the noise vector, ${\rm{DS}}_k$ is the desired signal, ${\rm {LS}}_k$ is the leaked signal, ${\rm {UI}}_{k,k'}$ represents the interference of the $k'$-th device, and ${\rm {N}}_k$ is the noise term. Then, the \(k\)-th user's achievable data rate can be expressed as
\begin{equation}
    \label{Ergodicrate}
    \small
    \begin{split}
        R^{\text{MRC}}_k &= B_i\log_2(1+\text{SINR}_k),\\
         \text{SINR}_k &= \frac{|\text{DS}_k|^2}{|\text{LS}_k|^2+\sum\limits_{k' \in \{\mathcal{K}_i \backslash k\}}|\text{UI}_{k,k'}|^2 +|\text{N}_k|^2},
    \end{split}
\end{equation}
where \(\text{SINR}_k\) is the signal-to-interference-plus-noise ratio (SINR) of the user \(k\).

\section{Performance Analysis and Problem Formulation}
\subsection{Performance Analysis}
As can be seen from (\ref{Ergodicrate}), it is challenging to derive the closed-form expression of the \(k\)-th user's ergodic data rate. To handle this issue, the lower bound of the $k$-th device's data rate is derived, which is given by
\begin{equation}
\setlength\abovedisplayskip{5pt}
\setlength\belowdisplayskip{5pt}
\label{kth_SINR}
\small
\begin{split}
    R^{\text{MRC}}_k &\ge {\underline R}_k^{\rm MRC} = B_i\log_2(1+\text{SINR}^{\text{LB}}_k)\\
    \text{SINR}^{\text{LB}}_k &= \frac{{{{p^d_k\left| \mathbb{E}\Big\{\sum\limits_{m \in \mathcal{M}_k} {w_{m,k}}\mathbf{\hat h}^H_{m,k}\mathbf{h}_{m,k}\Big\} \right|}^2}}}{{{ \mathbb{E}\left\{\left| \rm{LS}_k \right|^2 \right\} } + \sum\limits_{k'\in \{\mathcal{K}_i \backslash k\}} \mathbb{E}\{{{\left| {{\rm{UI}}_{k,k'}} \right|}^2}\}  + \mathbb{E}\{{\left| {{{\rm{N}}_k}} \right|}^2}\}}.
\end{split}
\end{equation}
In the following theorem, we derive a closed-form expression for the lower bound.

\begin{theorem}
\label{MRC_SINR_T}
The ergodic achievable rate for the $k$-th device using the MRC decoder can be lower bounded by
\begin{equation}
\setlength\abovedisplayskip{5pt}
\setlength\belowdisplayskip{5pt}
\small
\label{MRC_LB_rate}
{\underline R}_k^{\rm MRC} = B_i\log_2(1+\text{SINR}^{\text{LB}}_k),
\end{equation}
where $\text{SINR}^{\text{LB}}_k$ is given by (\ref{MRC_SINR_LB}) at the bottom of next page.

\begin{figure*}[b]
\hrulefill
\centering
\begin{equation}
\setlength\abovedisplayskip{5pt}
\setlength\belowdisplayskip{5pt}
\small
\label{MRC_SINR_LB}
\text{SINR}^{\text{LB}}_k = \frac{ p^d_k\left|\sum\limits_{m \in {\mathcal{M}}_k}{w_{m,k}}\Big\{ \tau p^p_k \text{tr}\{\mathbf{R}_{m,k}\boldsymbol{\Psi}_{m,k}\mathbf{R}_{m,k}\} + {\bar K}_{m,k}a_{m,k}||\mathbf{\bar{h} }_{m,k}||^2\Big\} \right|^2}{ I^{\text{noise}}_k  + \sum\limits_{k' \in \mathcal{K}_i}  p^d_{k'}I^{1}_{k,k'} +  \sum\limits_{k' \in \{\mathcal{K}_i \backslash k\}}  p^d_{k'}I^{2}_{k,k'}   + \sum \limits_{k' \in \{\mathcal{P}_k \backslash k \cap \mathcal{K}_i\} } p^d_{k'} I^{3}_{k,k'}},
\end{equation}
\end{figure*}

The terms of \( I^{\text{noise}} \), \(I^{1}_{k,k'}\), \(I^{2}_{k,k'}\), and \(I^{3}_{k,k'}\) in (\ref{MRC_SINR_LB})  are given by
\begin{equation}
    \label{noiseterm}
    \begin{split}
    \small
     I^{\text{noise}}_k& = \sum\limits_{m \in \mathcal{M}_k} w^2_{m,k}\Big( \tau p^p_k  \text{tr}\{\mathbf{R}_{m,k}\boldsymbol{\Psi}_{m,k}\mathbf{R}_{m,k}\} \\
     &+ \bar K_{m,k}a_{m,k} \mathbf{\bar h}^H_{m,k}\mathbf{\bar h}_{m,k}\Big)\sigma^2_i,   
    \end{split}
\end{equation}

\begin{equation}
    \begin{split}
    \small
        I^{1}_{k,k'} &=   \sum\limits_{m \in \mathcal{M}_k} w^2_{m,k}\bar K_{m,k'}\tau p^p_k a_{m,k'} \\
        &\times \mathbf{\bar h}^H_{m,k'} \mathbf{R}_{m,k}\boldsymbol{\Psi}_{m,k} \mathbf{R}_{m,k}\mathbf{h}_{m,k'} \\
        & + \sum\limits_{m \in \mathcal{M}_k} w^2_{m,k} \bar K_{m,k}a_{m,k}\mathbf{\bar h}^H_{m,k}\mathbf{R}_{m,k'} \mathbf{\bar h}_{m,k} \\
        & + \sum\limits_{m \in \mathcal{M}_k}  w^2_{m,k}\tau p^p_k \text{tr}\{\mathbf{R}_{m,k'}\mathbf{R}_{m,k}\boldsymbol{\Psi}_{m,k} \mathbf{R}_{m,k}\},
    \end{split}
\end{equation}

\begin{equation}
    \begin{split}
    \small
        I^{2}_{k,k'} &=    \left|\sum\limits_{m \in {\mathcal{M}}_k}{ w_{m,k}} \sqrt{\bar K_{m,k}\bar K_{m,k'}a_{m,k}a_{m,k'}}\mathbf{\bar h}^H_{m,k}\mathbf{\bar h}_{m,k'}\right|^2 ,
    \end{split}
\end{equation}
and 
\begin{equation}
    \begin{split}
    \small
        I^{3}_{k,k'} &=   2 \tau\sqrt{ p^p_{k'}p^p_{k}} \text{Re}\Big\{\sum\limits_{n \in \mathcal{M}_k}w_{n,k}\text{tr}\{\mathbf{R}_{n,k}\boldsymbol{\Psi}_{n,k}\mathbf{R}_{n,k'}\} \\
        &\times \sum\limits_{m \in \mathcal{M}_k} w_{m,k} \sqrt{ \bar K_{m,k}a_{m,k}\bar K_{m,k'}a_{m,k'} }  \mathbf{\bar h}^H_{m,k}\mathbf{\bar{h} }_{m,k'} \Big\} \\
        & + \tau^2 p^p_k p^p_{k'}\Big(\sum\limits_{m \in \mathcal{M}_k}{w_{m,k}} \text{tr}\{\mathbf{R}_{m,k}\boldsymbol{\Psi}_{m,k}\mathbf{R}_{m,k'} \}  \Big)^2.
    \end{split}
\end{equation}

\emph{Proof}: Please refer to Appendix \ref{Prooftheorem1}. $\hfill\blacksquare$

\end{theorem}

As can be seen from (\ref{MRC_SINR_LB}), SINR is affected by the pilot power \(p^p_k\), the transmission power \(p^d_k\), the Rician factor \(\bar K_{m,k}\), and the pilot sharing set \(\mathcal{P}_k\). To further characterize the impacts of these factors on the data rate, a comprehensive analysis is given in the following.

\begin{corollary}
Considering the LoS channel, i.e., \(\bar{K}_{m,k} \rightarrow \infty\), we have \(a_{m,k} \rightarrow 0\), \(\mathbf{R}_{m,k}\boldsymbol{\Psi}_{m,k}\mathbf{R}_{m,k}\ \rightarrow \mathbf{0}_N\), and \(\bar{K}_{m,k} a_{m,k} \rightarrow \beta_{m,k}\). Then, the \(k\)-th user's SINR can be lower bounded by (\ref{LosSINR}), which is given at the bottom of the next page.

\begin{figure*}[b]
\hrulefill
   \begin{equation}
\label{LosSINR}
    \begin{split}
    \small
        \text{SINR}^{\text{LoS}}_k = \frac{p^d_k N^2 (\sum\limits_{m \in \mathcal{M}_k}{w_{m,k}} \beta_{m,k})^2}{\sum\limits_{m \in \mathcal{M}_k}w^2_{m,k}N \sigma^2_i \beta_{m,k}+ \sum\limits_{k'  \in \{\mathcal{K}_i \backslash k\}} p^d_{k'}\left| \sum\limits_{m \in \mathcal{M}_k  }w_{m,k}\sqrt{\beta_{m,k}\beta_{m,k'}}\mathbf{\bar h}^H_{m,k}\mathbf{\bar h}_{m,k'}\right|^2 }.
    \end{split}
\end{equation} 
\end{figure*}

\end{corollary}
As can be seen from (\ref{LosSINR}), there is no interference caused by pilot contamination and uncertain beamforming errors when considering the pure LoS channel. Furthermore, it's worth noting that severe inter-user interference can cause performance loss. Therefore, it is worth investigating the combination of strategy, power control, and user scheduling schemes to maximize the sum data rate of satellite networks.

\subsection{Problem Formulation}
Based on the derived lower bound given in (\ref{MRC_LB_rate}), we aim to maximize the sum data rate by optimizing the user scheduling strategy \(\mathcal{K}_i\), bandwidth allocation \(B_i\), transmission power \(p^d_k\), and weights \(w_{m,k}\), \(\forall m,k\), while satisfying each user's transmission requirements \footnote{We assume that the pilot power of each user is fixed and random pilot allocation is adopted.}. The problem can be equivalently written as 
\begin{subequations}
\setlength\abovedisplayskip{5pt}
\setlength\belowdisplayskip{5pt}
\label{MRC_optimization}
\begin{align}
\small
\mathop {\max }\limits_{\left\{ {p_k^d} \right\},\left\{w_{m,k}\right\}, \left\{ \mathcal{K}_i\right\},\left\{ {B}_i\right\}} \quad & \sum\limits_{i=1}^{I}\sum\limits_{k \in \mathcal{K}_i} B_i\log_2(1+\text{SINR}^{\text{LB}}_k)\notag\\
{\rm{s}}{\rm{.t}}{\rm{.}}\;\;\;\; & {\underline R}_k^{\rm MRC} \ge R_k^{{\rm{req}}},\forall k \in \mathcal{K}_i, \forall i  \label{MRC_optimization_b}\\
& {p^d_k \le P^{d,\max}_{k},\forall k}, \label{MRC_optimization_c} \\
& \sum\limits_{m \in \mathcal{M}_k} w^2_{m,k} = 1, \forall k, \label{MRC_optimization_d}\\
& \sum\limits_{i=1}^I B_i \leq B, \label{MRC_optimization_e} \\
& \sum\limits_{i=1}^I |\mathcal{K}_i| = K, \label{MRC_optimization_f} \\
&  0 < |\mathcal{K}_i| \leq N_{\max}, \forall i, \label{MRC_optimization_H} \\
& \mathcal{K}_i \cap \mathcal{K}_j = \emptyset, i \neq j, \label{MRC_optimization_g} 
\end{align}
\end{subequations}
where \(|\mathcal{K}_i|\) is the cardinality of \(\mathcal{K}_i\), \(N_{\max}\) is the maximum number of user in the \(i\)-th sub-band, constraint (\ref{MRC_optimization_b}) denotes the minimum data rate \(R_k^{\text{Req}}\) of the user \(k\), constraint (\ref{MRC_optimization_c}) means that each device's transmission should be less than the maximum power \(p_k^{d, \max}\), and constraint (\ref{MRC_optimization_d}) implies that the normalized power of the detected signal is equal to 1. Constraint (\ref{MRC_optimization_e}) means the limited bandwidth, constraint (\ref{MRC_optimization_f}) implies that all users will be served simultaneously, constraint (\ref{MRC_optimization_H}) ensures the fairness of each sub-band, and constraint (\ref{MRC_optimization_g}) denotes that any two sets are non-overlapping.  

As can be seen from Problem (\ref{MRC_optimization}), it is an NP-hard problem, and thus finding the optimal solution is demanding. Furthermore, owing to the constraints (\ref{MRC_optimization_g}) and (\ref{MRC_optimization_f}), the intractable form make the problem more challenging. Therefore, we aim to proposed an efficient algorithm to find the sub-optimal solution of Problem (\ref{MRC_optimization}).

\section{Proposed Algorithm}
To deal with this difficulty, we decompose it into three problems, namely, user scheduling strategy, bandwidth allocation, and joint combination and power control. Then, we alternatively optimize these variables by using the graph coloring theory, the SCA method, and GP.

\subsection{User Scheduling Strategy}
With the fixed bandwidth, transmission power, and combination strategy, Problem (\ref{MRC_optimization}) can be simplified to
\begin{subequations}
\setlength\abovedisplayskip{5pt}
\setlength\belowdisplayskip{5pt}
\label{UG_optimization}
\begin{align}
\small
\mathop {\max }\limits_{\left\{ \mathcal{K}_i\right\}} \quad & \sum\limits_{i=1}^{I}\sum\limits_{k \in \mathcal{K}_i}  B_i\log_2(1+\text{SINR}^{\text{LB}}_k), \notag\\
{\rm{s}}{\rm{.t}}{\rm{.}}\;\;\;\; & \text{SINR}^{\text{LB}}_k \ge 2^{\frac{R_k^{{\rm{req}}}}{B_i}} -1,\forall k \in \mathcal{K}_i, \forall i,\label{UG_optimization_a}\\
 & (\ref{MRC_optimization_f}), (\ref{MRC_optimization_H}), (\ref{MRC_optimization_g}). \label{UG_optimization_b}
\end{align}
\end{subequations}
As can be seen from (\ref{UG_optimization}), it is a resource-assignment problem and thus still challenging to obtain the optimal solution in general. The only way to obtain the optimal solution is by using the exhaustive search method, the complexity of which increases exponentially with the number of UEs and sub-bands. To tackle this issue, we aim to transform it into a more tractable form and then provide a low-complexity algorithm.

\begin{figure*}
	\centering
	\subfigure[Undirected graph based on the correlation factors \(\rho_{k,k'}\).]{
		\begin{minipage}[t]{0.3\linewidth}
			\centering
			\includegraphics[width=2in]{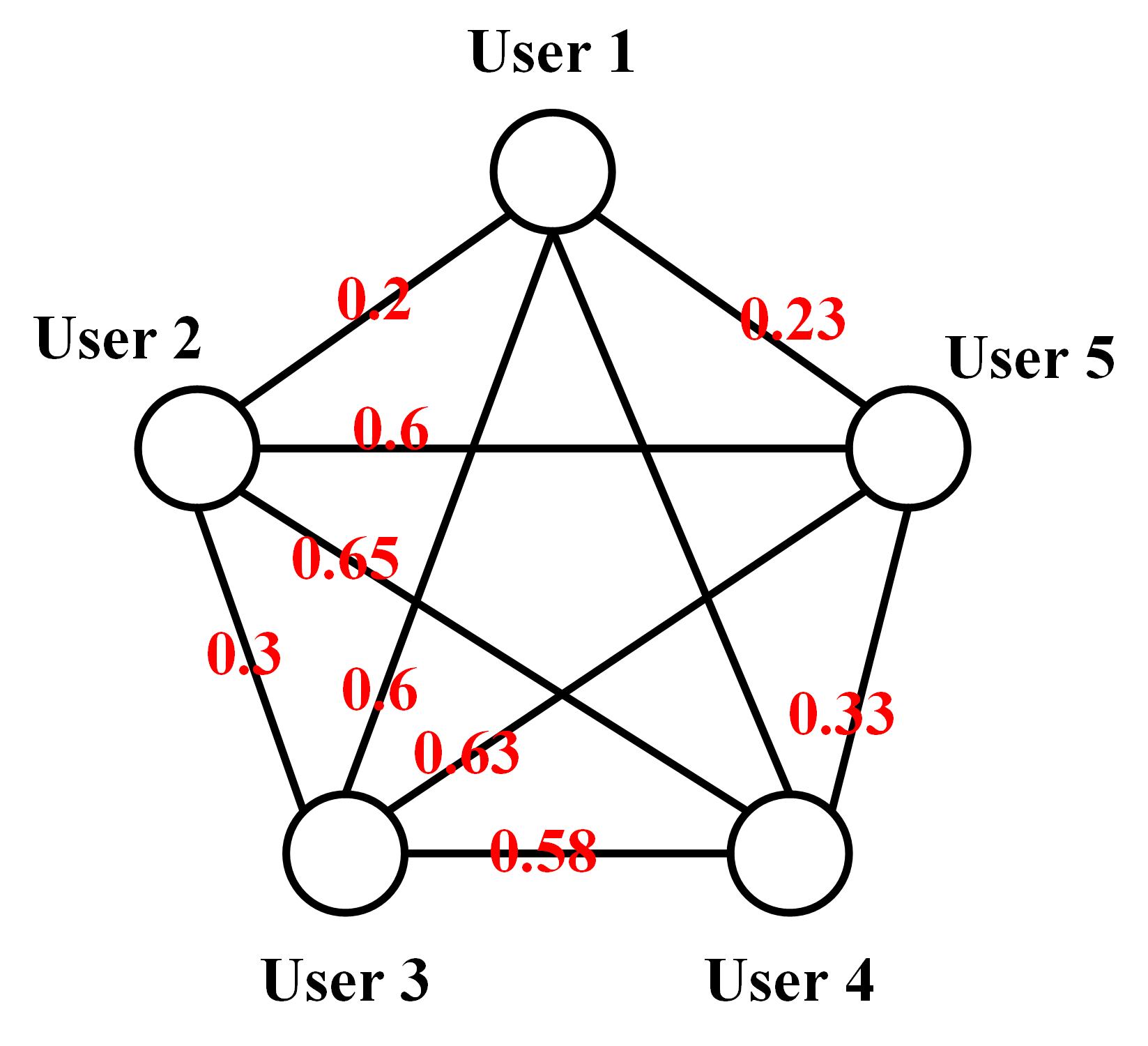}\hspace{10mm}
		\end{minipage}}
	\quad
	\subfigure[Sub-band assignment based on the Dsatur algorithm.]{\begin{minipage}[t]{0.3\linewidth}
			\centering
			\includegraphics[width=2in]{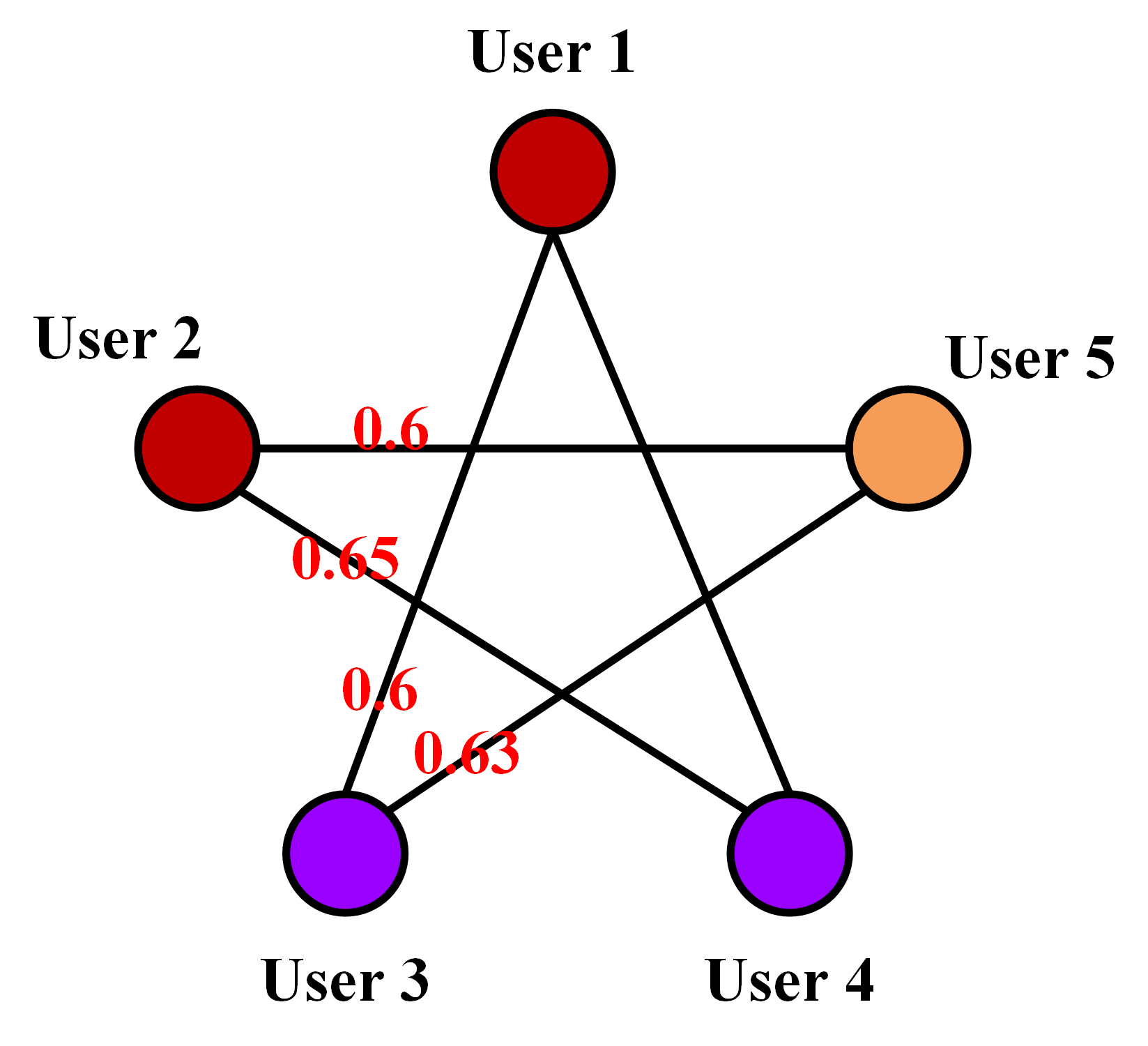}\hspace{10mm}
		\end{minipage}}
	\quad
	\subfigure[Sub-band allocation after considering minimal requirements.]{\begin{minipage}[t]{0.3\linewidth}
			\centering
			\includegraphics[width=2in]{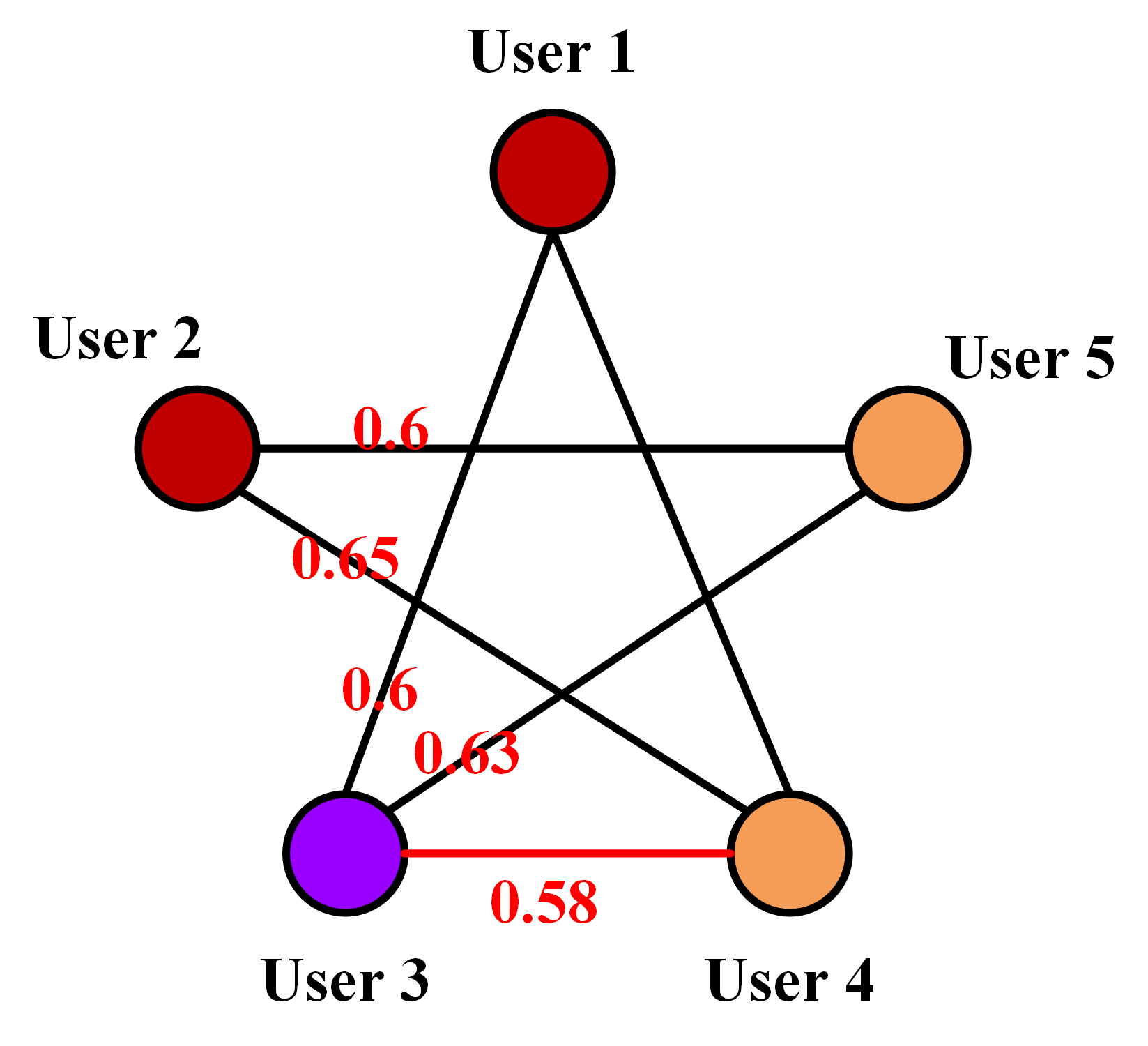} \hspace{10mm}
	\end{minipage}}
	\caption{Undirected graph and sub-band allocation schemes.}
	\label{cluster}
\end{figure*}

Firstly, it can be observed from (\ref{MRC_SINR_LB}) that the dominant interference comes from the user's interference from the LoS channel and pilot contamination, owing to the correlation of channels. Therefore, the correlated factor \(\rho_{k,k'}\) between the \(k\)-th and the \(k'\)-th users can be defined as
\begin{equation}
    \label{increase}
    \small
    \begin{split}
        \rho_{k,k'} = \frac{\sum\limits_{m\in \mathcal{M}_k} \mathbf{\hat h}^H_{m,k}\mathbf{\hat h}_{m,k'} }{\sum\limits_{m\in \mathcal{M}_k} \mathbf{\hat h}^H_{m,k} \mathbf{\hat h}_{m,k}} + \frac{\sum\limits_{m\in \mathcal{M}_{k'}} \mathbf{\hat h}^H_{m,k'}\mathbf{\hat h}_{m,k} }{\sum\limits_{m\in \mathcal{M}_{k'}} \mathbf{\hat h}^H_{m,k'} \mathbf{\hat h}_{m,k'}}.
    \end{split}
\end{equation}
Based on the given correlated factor \(\rho_{k,k'}\), we can readily construct a binary matrix \(\mathbf{B} \in \mathbb{N}^{K \times K}\) to illustrate whether the \(k\)-th user and the \(k'\)-th user can share the common sub-band or not. The element \([\mathbf{B}]_{k,k'} \) can be defined as
\begin{equation}
\small
    [\mathbf{B}]_{k,k'} =\left\{ \begin{array}{*{20}{c}}
       {1,}  & {\text{if} \; \rho_{k,k'} \ge \rho_{\text{Th}}},  \\
        {0,} & {\text{otherwise},}
    \end{array} \right.
\end{equation}
where \(x\) represents \(k\) or \(k'\) and \(\rho_{\text{th}}\) is the threshold. If the correlated factor \([\mathbf{B}]_{k,k'} \) is 1, then sharing the common sub-band between the \(k\)-th user and the \(k'\)-th user will cause severe interference. Conversely, the \(k'\)-th user is the potential candidate for sharing the same sub-band with the user \(k\), when \([\mathbf{B}]_{k,k'} = 0\). Therefore, Problem (\ref{UG_optimization}) can be transformed into a coloring graph, which can be readily solved by using the DSatur algorithm \cite{san2012new,peng2023pilotsharing}. However, the DSatur algorithm focuses on the minimum number of colors \(n_c\), ignoring each user's minimum requirement. To elaborate on this issue, the user scheduling strategy can be illustrated in Fig. \ref{cluster}. The correlation among all users is shown in Fig. \ref{cluster}(a), where the red number represents the correlated factor \(\rho_{k,k'}\). Then, by setting the threshold \(\rho_{\text{Th}}\) to 0.6, the sub-band allocation based on the DSatur algorithm is depicted in Fig. \ref{cluster}(b), where the same color indicates that the same sub-band is used. As can be seen, sharing the purple color between user 3 and user 4 may lead to severe interference, thereby violating the requirement of each user.

To solve this challenge, an iterative algorithm is proposed to modify the allocation strategy, which is detailed in Algorithm 1. By linking user 3 and user 4 by the red line, the updated allocation strategy is illustrated in Fig. \ref{cluster}(c). Specifically, the iteration number \(l\) is initialized, and the correlation factor \(\{\rho^{(l)}_{k,k'}\}\) can be calculated based on (\ref{increase}). The threshold \(\rho^{(l)}_{\text{th}}\) is initialized as the averaged value of \(\{\rho^{(l)}_{k,k'}\}\), and the binary matrix \(\mathbf{B}^{(l)}\) is constructed without considering the minimum requirement of data rate. Then, using the DSatur algorithm to obtain the allocation strategy \(\{\mathcal{K}^{(l)}_i\}\) with \(n^{(l)}_c\). After applying the DSatur algorithm, two cases arise: \(n_c > I\) and \(n_c \le I\). In the case of  \(n_c > I\), this implies that more available sub-bands are needed. When \(n_c \le I\), the available sub-bands can be reused or reassigned, and further adjustments may be applied. In the following, we discuss the two cases in detail.
\begin{itemize}
    \item If \(n_c > I\), the threshold \(\rho^{(l)}_{\text{Th}}\) can be updated as \(\rho^{(l)}_{\text{Th}} = \frac{\rho^{(l-1)}_{\text{Th}} + \max\{\rho_{k,k'}\}}{2}\). Then, the binary matrix \(\mathbf{B}^{(l)}\) without considering constraint (\ref{UG_optimization_a}) can be calculated accordingly, and the \(l\)-th allocation strategy \(\{\mathcal{K}^{(l)}_i\}\) can be obtained by using the DSatur algorithm.
    \item For \(n_c \le I\), the threshold \(\rho^{(l)}_{\text{Th}}\) can be updated as \(\rho^{(l)}_{\text{Th}} = \rho^{(l-1)}_{\text{Th}} \). Then, we update \(\mathbf{B}^{(l)}\) while considering the minimum requirement of data rate. Specifically, the users with poor performance can be found by \(\mathcal{S}^{(l)} = \{k| \mathop{\arg\min}\limits_{k} \text{SINR}^{\text{LB}}_k, k \in \mathcal{K}\}\), and then the users with the largest interference can be obtained by \(\mathcal{C}^{(l)} = \{k'| = \mathop{\arg\min}\limits_{k'} \text{SINR}^{\text{LB}}_{k'}, k' \in \{\mathcal{K}_i \backslash k\}\). After that, the binary matrix can be updated by setting \(b_{k,k'} = 1\), \(k \in \mathcal{S}^{(l)}\), \(k' \in \mathcal{C}^{(l)}\) and the scheduling strategy \(\{\mathcal{K}_i^{(l)}\}\) can be obtained.
\end{itemize}
Finally, if the constraints of Problem (\ref{UG_optimization}) are satisfied and the sum data rate is larger than that based on the previous strategy, we update the scheme and optimal sum data rate. Otherwise, we continue to update the scheduling strategy until constraints are satisfied or the iteration number is larger than \(L_{\max}\) \footnote{If the iteration number \(l > L_{\max}\), then some users may be scheduled next time owing to severe interference.}.

\begin{algorithm}[t]
	\caption{Iterative Algorithm For Solving Problem (\ref{UG_optimization})}
	\begin{algorithmic}[1]
		\label{UGsolving}
                    \STATE Initialize the iterative number \(l = 0\), maximum iteration number \(L_{\max} = 100\), and the optimal user scheduling strategy \(\{\mathcal{K}^{\text{opt}}_i\}\) with sum data rate \({S}^{\text{opt}}_{d}\);
                    \STATE Calculate the correlation \(\{\rho^{(l)}_{k,k'}\}\) by using (\ref{increase}) and the threshold \(\rho^{(l)}_{\text{Th}}\) by taking the averaged correlation factor. Then, construct the binary matrix \(\mathbf{B}^{(l)}\) and obtain the allocation strategy \(\{\mathcal{K}^{(l)}_i\}\) with \(n^{(l)}_c\);
                    \WHILE{\(l \le L_{\max}\)}
                    \STATE Update iterative number \(l = l + 1\);
                    \IF{\( n^{(l-1)}_c > I\)}
                    \STATE Calculate \(\rho^{(l)}_{\text{Th}} =\frac{\rho^{(l-1)}_{\text{Th}} + \max\{\rho_{k,k'}\}}{2} \), and then update the binary matrix \(\mathbf{B}^{(l)}\) and allocation strategy \(\{\mathcal{K}^{(l)}_i\}\) with \(n^{(l)}_c\) accordingly.
                    \ELSIF{\( n^{(l-1)}_c \le I \)}
                    \STATE Update  \(\rho_{\text{Th}}^{(l)} = \rho_{\text{Th}}^{(l-1)}\);
                    \STATE Find the users with smallest data rate, i.e., \(\mathcal{S}^{(l)} = \{k = \mathop{\arg\min}\limits_{k} \text{SINR}^{\text{LB}}_k, k \in \mathcal{K}\}\), and then find the user with the largest interference by \(\mathcal{C}^{(l)} = \{k'| = \mathop{\arg\min}\limits_{k'} \text{SINR}^{\text{LB}}_{k'}, k' \in \{\mathcal{K}_i \backslash k\}\}\);
                    \STATE Update the binary matrix \(\mathbf{B}^{(l)}\) by using \(b_{k,k'} = 1\), \(k \in \mathcal{S}^{(l)}\), \(k' \in \mathcal{C}^{(l)}\), obtain the scheduling strategy \(\{\mathcal{K}^{(l)}_i\}\) and the number of colors \(n_c^{(l)}\) by adopting the DSatur algorithm, and calculate the sum data rate \(S^{(l)}_d\);
                    \ENDIF
                    \IF{All constraints of Problem (\ref{UG_optimization}) are satisfied}
                    \IF{\(S^{(l)}_d \ge S^{\text{opt}}_{d}\)}
                    \STATE Update the optimal user scheduling strategy \(\mathcal{K}^{\text{opt}}_i = \mathcal{K}_i\), \(\forall i\), and update the optimal sum data rate with \(S^{\text{opt}}_{d} =S^{(l)}_d  \)  ;
                    \ENDIF
                    \ENDIF
                    \ENDWHILE
	\end{algorithmic}
\end{algorithm}



\subsection{Joint Combination and Power Control}
With given bandwidth allocation and fixed sub-bands, we introduce the auxiliary variables \(\chi_k\), \(\forall k\) and equivalently transform Problem (\ref{MRC_optimization}) into
\begin{subequations}
\setlength\abovedisplayskip{5pt}
\setlength\belowdisplayskip{5pt}
\label{AO}
\begin{align}
\small
\mathop {\max }\limits_{\left\{\chi_k\right\},\left\{ {p_k^d} \right\},\left\{w_{m,k}\right\}} \quad & \sum\limits_{i = 1}^{I}\sum\limits_{k \in \mathcal{K}_i}B_i \log_2(1+\chi_k) \notag\\
{\rm{s}}{\rm{.t}}{\rm{.}}\;\;\;\; &  \chi_k \ge 2^{\frac{R_k^{{\rm{req}}}}{B_i}} -1,\forall k,  \label{AO_a}\\
& \text{SINR}^{\text{LB}}_k   \ge \chi_k, \forall k, \label{AO_b}\\
& (\ref{MRC_optimization_c}),(\ref{MRC_optimization_d}).\label{AO_c} 
\end{align}
\end{subequations}

Owing to the Lemma 2 given in \cite{peng2022resource}, the objective function in (\ref{AO}) can be iteratively approximated as
\begin{equation}
    \label{appro}
    \small
    \log_2(1+\chi_k) \ge \psi_k \log_2(\chi_k) + \delta_k.
\end{equation}
\(\psi_k\) and \(\delta_k\) can be given by
\begin{equation}
    \label{rho}
    \small
    \psi_k = \frac{\chi^{(l)}_k}{1+\chi^{(l)}_k},
\end{equation}
and 
\begin{equation}
    \label{delta}
    \small
    \delta_k = \log_2(1+\chi^{(l)}_k) -\frac{\chi^{(l)}_k}{1+\chi^{(l)}_k} \log_2(\chi^{(l)}_k),
\end{equation}
where \(\chi^{(l)}_k\) is the \(l\)-th iteration of \(\chi_k\).

Then, by ignoring the constant term in the objective function, Problem (\ref{AO}) can be transformed into
\begin{subequations}
\setlength\abovedisplayskip{5pt}
\setlength\belowdisplayskip{5pt}
\label{AOa}
\begin{align}
\small
\mathop {\max }\limits_{\left\{\chi_k\right\},\left\{ {p_k^d} \right\},\left\{w_{m,k}\right\}}  \quad &\prod\limits_{k \in \mathcal{K}} (\chi_k)^{{\hat \psi_k}^{(l)}}\notag\\
{\rm{s}}{\rm{.t}}{\rm{.}}\;\;\;\; & (\ref{AO_a}),(\ref{AO_b}),(\ref{MRC_optimization_c}),(\ref{MRC_optimization_d}),\label{AOa_a}
\end{align}
\end{subequations}
where \({\hat \psi_k}^{(l)}\) is \(\psi_k B_i\) in the \(l\)-th iteration. However, owing to the complicated form of (\ref{AO_b}), this is not a GP problem. To tackle this issue, we have the following lemma.

\begin{lemma}
\label{lemma1}
For the determined constant \(A_{m,k} > 0\) and given weight \({\hat w}_{m,k} > 0 \), \(\forall m,k\), \(\Big(\sum\limits_{m \in \mathcal{M}_k} {w_{m,k} }A_{m,k}\Big)^2\) is lower bounded by
\begin{equation}
    \label{lower}
    \begin{split}
    \small
        \Big(\sum\limits_{m \in \mathcal{M}_k} {w_{m,k} }A_{m,k}\Big)^2 \ge c_k\prod\limits_{m \in \mathcal{M}_k}(w_{m,k})^{a_{m,k}},
    \end{split}
\end{equation}
where \(a_{m,k}\) and \(c_k\) can be expressed as
\begin{equation}
    \label{amk}
    \small
    a_{m,k} = \frac{{{\hat w}_{m,k}}A_{m,k}}{\sum\limits_{n \in \mathcal{M}_k}{{\hat w}_{n,k}}A_{n,k}},
\end{equation}
and 
\begin{equation}
    \label{ck}
    \small
    c_k = \frac{\Big(\sum\limits_{m \in \mathcal{M}_k} {{\hat w}_{m,k} }A_{m,k}\Big)^2}{\prod\limits_{m \in \mathcal{M}_k}({\hat w}_{m,k})^{a_{m,k}}}. 
\end{equation}
The inequality holds only when \(w_{m,k}={\hat w}_{m,k}\), \(\forall m, k\).

\emph{Proof}: The proof is omitted owing to the similar process of Appendix F in \cite{peng2023resource}. \(\hfill\blacksquare\)
\end{lemma}

By using Lemma \ref{lemma1}, Problem (\ref{AOa}) can be written as
\begin{subequations}
\setlength\abovedisplayskip{5pt}
\setlength\belowdisplayskip{5pt}
\label{AOb}
\begin{align}
\small
\mathop {\max }\limits_{\left\{\chi_k\right\},\left\{ {p_k^d} \right\},\left\{w_{m,k}\right\}}  \quad &\prod\limits_{k \in \mathcal{K}} (\chi_k)^{{\hat \psi_k}^{(l)}}\notag\\
{\rm{s}}{\rm{.t}}{\rm{.}}\;\;\;\; & p^d_kc^{(l)}_k\prod\limits_{m \in \mathcal{M}_k}(w_{m,k})^{a^{(l)}_{m,k}}  \ge \chi_k \times  \notag \\
&\Big( \sum\limits_{k' \in \mathcal{K}_i}  p^d_{k'}I^{1}_{k,k'}  +  \sum\limits_{k' \in \{\mathcal{K}_i \backslash k\}}  p^d_{k'}I^{2}_{k,k'} \notag \\
& + I^{\text{noise}}_k + \sum \limits_{k' \in \{\mathcal{P}_k \backslash k \cap \mathcal{K}_i\} } p^d_{k'} I^{3}_{k,k'} \Big),  \forall k, \\
& (\ref{AO_a}),(\ref{MRC_optimization_c}),(\ref{MRC_optimization_d}),\label{AOb_a}
\end{align}
\end{subequations}
where \(c^{(l)}\) and \(a^{(l)}_{m,k}\) are the \(c_k\) and \(a_{m,k}\) in the \(l\)-th iteration, respectively. Note that this is a GP problem that can be readily solved by using CVX. Furthermore, the feasible region can be found by solving the following problem, which can be expressed as
\begin{subequations}
\setlength\abovedisplayskip{5pt}
\setlength\belowdisplayskip{5pt}
\label{AOc}
\begin{align}
\small
\mathop {\max }\limits_{\left\{ {p_k^d} \right\},\left\{w_{m,k}\right\},\phi}  \quad &\phi\notag\\
{\rm{s}}{\rm{.t}}{\rm{.}}\;\;\;\; & p^d_kc^{(l)}_k\prod\limits_{m \in \mathcal{M}_k}(w_{m,k})^{a^{(l)}_{m,k}}  \ge {\phi(2^{\frac{R_k^{{\rm{req}}}}{B_i}} -1)} \times \notag \\
&\Big(\sum\limits_{k' \in \mathcal{K}_i}  p^d_{k'}I^{1}_{k,k'} +  \sum\limits_{k' \in \{\mathcal{K}_i \backslash k\}}  p^d_{k'}I^{2}_{k,k'}\notag \\
& +I^{\text{noise}}_k   + \sum \limits_{k' \in \{\mathcal{P}_k \backslash k \cap \mathcal{K}_i\}} p^d_{k'} I^{3}_{k,k'} \Big), \forall k, \\
& (\ref{MRC_optimization_c}),(\ref{MRC_optimization_d}).\label{AOc_a}
\end{align}
\end{subequations}
Obviously, Problem (\ref{AOc}) can be readily solved by using CVX. Furthermore, \(\phi \ge 1\) implies that the feasible region is found.

Based on the above discussions, Problem (\ref{AO}) can be solved iteratively, which is detailed in Algorithm \ref{FP2}.

\begin{algorithm}[t]
	\caption{Alternating Iterative Algorithm For Solving Problem (\ref{AO})}
	\begin{algorithmic}[1]
		\label{FP2}
                    \STATE Initialize the iterative number \(l = 1\), the error tolerance \(\epsilon = 0.01\).
                    \STATE Initialize the transmission power \(\{p^{d,(l)}_k = P^{d,\max}_k\}\) and weight \(\{w^{(l)}_{m,k}\}\), \(\forall k,m\), by solving Problem (\ref{AOc});
                    \STATE Calculate the sum data rate of \(K\) users, denoted as \(\text{Obj}^{(l)} \), and define the initial objective function \(\text{Obj}^{(0)} = 0\);
                    \WHILE{\(\frac{\text{Obj}^{(l)} -\text{Obj}^{(l-1)}}{\text{Obj}^{(l)}} \ge \epsilon   \)}
                    \STATE Update \(\{\hat \psi^{(l)}_k\}\);
                    \STATE Update iterative number \(l = l + 1\);
                    \STATE Using CVX to solve Problem (\ref{AOb}) obtain the transmission power \(\{p^{d,(l)}_k\}\) and weight \(\{w^{(l)}_{m,k}\}\);
                    \STATE Calculate the objective function \(\text{Obj}^{(l)} \);
                    \ENDWHILE
	\end{algorithmic}
\end{algorithm}

\subsection{Bandwidth Allocation}
For the given user scheduling strategy, transmission power, and combination scheme, Problem (\ref{MRC_optimization}) can be rewritten as
\begin{subequations}
\setlength\abovedisplayskip{5pt}
\setlength\belowdisplayskip{5pt}
\label{bandwidth}
\begin{align}
\small
\mathop {\max }\limits_{\left\{ {B}_i\right\}} \quad & \sum\limits_{i=1}^{I}\sum\limits_{k \in \mathcal{K}_i} B_i\log_2(1+\text{SINR}^{\text{LB}}_k)\notag\\
{\rm{s}}{\rm{.t}}{\rm{.}}\;\;\;\; & {\underline R}_k^{\rm MRC} \ge R_k^{{\rm{req}}},\forall k \in \mathcal{K}_i, \forall i  \label{bandwidth_b}\\
& \sum\limits_{i=1}^I B_i \leq B. \label{bandwidth_c}
\end{align}
\end{subequations}
Owing to its intractable form, it is challenging to recognize the convexity of Problem (\ref{bandwidth}). To tackle this issue, each user's data rate can be defined as \(f(x) = x \log_2(1+\frac{a}{bx + c})\) with the given sub-bands allocation, power control, and weights, where \(a\), \(b\) , \(c\) , and \(x\) mean the desired power, noise power related bandwidth, interference plus leaked power, and bandwidth, respectively. By deriving the first-order and second-order derivatives of \(f(x)\), we have 
\begin{equation}
    \small
    \begin{split}
    f'(x) &= \log_2(1+\frac{a}{bx + c}) +x \times \frac{bx+c}{bx+c+a}\times \frac{-abx}{(bx+c)^2\ln2}\\
    & = \log_2(1+\frac{a}{bx + c})-\frac{abx}{\ln2(bx+c)(bx+c+a)}, 
    \end{split}
\end{equation}
and 
\begin{equation}
    \small
    \begin{split}
    f''(x) &=-\frac{ab}{\ln2(bx+c)(bx+c+a)}\\
    &-\frac{ab(bx+c)(bx+c+a)-ab^2x(2bx+2c+a)}{\ln2(bx+c)^2(bx+c+a)^2}\\
    & = \frac{-ab(abx+2bcx+2c^2+2ac)}{\ln2(bx+c)^2(bx+c+a)^2} <0.
    \end{split}
\end{equation}
Finally, we prove that \(f(x)\) is a concave function with respect to \(x\) owing to \(a,b,c > 0\). Therefore, Problem (\ref{bandwidth}) can be readily solved by using CVX.

Based on the above-mentioned, we iteratively optimize the sub-band allocation, joint weights and power control, and bandwidth allocation to maximize the sum data rate of satellite networks.

\subsection{Algorithm Analysis}
As the user scheduling strategy with a larger sum data rate can be obtained, it is readily to prove the convergence of Algorithm \ref{UGsolving}. Since the complexity of Algorithm \ref{UGsolving} relies on the number of iterations \(N_{\text{A1}}\) and the complexity of the DSatur algorithm \(\mathcal{O}(K^2)\), the total complexity of Algorithm \ref{UGsolving} is \(\mathcal{O}(N_{\text{A1}}K^2)\). For Algorithm \ref{FP2}, the convergence can be proved by using a similar process in \cite{peng2023resource}, and thus it's omitted here for brevity. The complexity of Algorithm \ref{FP2} mainly depends on the complexity of each iteration and the number of iterations \(N_{\text{A2}}\). For each iteration, the computational complexity of this algorithm is ${\mathcal{O}}(N_{\text{A2}} \times \max\{(4K)^{3}), N_{f}\})$, where $N_{f}$ is the computational complexity of calculating the first-order and second-order derivatives of the objective function and constraint functions of Problem (\ref{FP2}) \cite{van2018joint}. The convergence for solving Problem (\ref{bandwidth}) can be readily proved, owing to the concave function, which is omitted for brevity. Furthermore, the complexity for solving Problem (\ref{bandwidth}) is \(\mathcal{O}(N_{\text{A3}}I^3)\), where \(N_{\text{A3}}\) is the number of iterations.  Based on these discussions, the total complexity for solving Problem (\ref{MRC_optimization}) is \(\mathcal{O}(N_{\text{A1}}K^2 + N_{\text{A2}} \times \max\{(4K)^{3}), N_{f}\} + N_{\text{A3}}I^3)\).
Finally, we will show that the proposed algorithm can achieve a locally optimal solution with rapid convergence.

\section{Simulation Results} 
This section presents numerical results to validate the derived lower bound based on MRC detection and demonstrate the performance improvement achieved by our proposed scheme. 
\subsection{Simulation Setup}
In this section, we assume that the satellite constellation follows the distribution of Starlink, while \(K\) users are randomly positioned on the Earth's surface with elevation angles ranging from 20 to 20.1 degrees. The free space loss is related to the carrier frequency \(f_c\), the distance between user and satellite \(d_{is}\), the speed of light \(c_s\), the transmission gain \(G_t\), and the receiving gain \(G_r\), which is given by \(20\lg(\frac{4\pi d_{is} f_c}{c_s}) - G_r - G_t\) (dB). Furthermore, the Rician factors are generated based on Table 6.7.2-1a in \cite{3GPP}, and the small-scale fading is generally modeled as Rayleigh fading with zero mean and unit variance. The noise power is related to the bandwidth \(B\), Boltzmann constant \(k_B\), noise temperature \(T_0\), and noise figure \(N_{\rm{dB}}\), and is denoted as \(\sigma^2_i = B_i \times {k_B} \times {T_0} \times  10^{\frac{N_{\rm{dB}}}{10}} \left( {\rm{W}} \right)\). Unless otherwise specified, the simulation parameters are summarized in Table \ref{tab: Margin_settings}. Additionally, based on the user-centric approach \cite{peng2023pilotsharing}, we choose the satellites based on the large-scale fading parameters, which are arranged in descending order.


\begin{table}[t]
        \small
        \caption{Simulation Parameters}
		\centering
		\begin{tabular}{|c|c|}\hline
	    Parameters Setting & Value  \\ \hline
        Carrier frequency ($f$) & 2 GHz  \\
        Bandwidth ($B$) & 1 MHz \\
        
		Transmission gain (\(G_t\)) & 0 dBi \\
        Receiving gain (\(G_r\)) & 6 dBi \\
        Boltzmann constant (\(k_B\)) & \(1.381 \times 10^{-23}\) \\
        Noise temperature \(T_0\) & 290 \\
        Noise figure ($N_{\rm{dB}}$) & 9 dB \\
        Number of antennas ($N$) & 100\\ \hline
		\end{tabular}
		\label{tab: Margin_settings}
\end{table}

\subsection{Channel Estimation}

\begin{figure}
    \centering
    \includegraphics[width=0.8\linewidth]{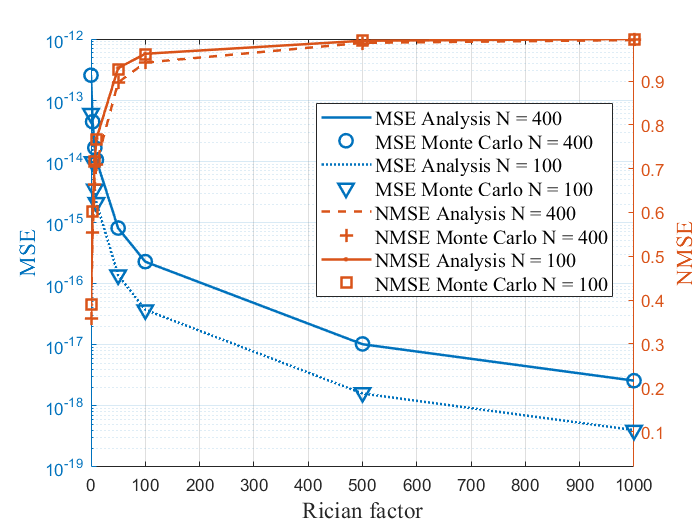}
    \caption{Monte Carlo Simulation vs. derived MSE of user 1 with \(K = 5\) and \(\tau = 3\).}
    \label{MSE}
\end{figure}
As shown in Fig. \ref{MSE}, we investigate the channel estimation by assuming the perfect LoS channel. As observed, the MSE decreases with an increasing Rician factor, while the NMSE exhibits the opposite trend. This is due to the fact that the power nLoS channel is decreasing with a larger Rician factor, leading to a smaller MSE. Meanwhile, the negligible nLoS channel cannot be estimated from noise, and thus NMSE tends to be 1 when the Rician factor goes to infinity, which validates our analysis in Remark 1.

\subsection{Derived Lower Bound}

\begin{figure}
    \centering
    \includegraphics[width=0.8\linewidth]{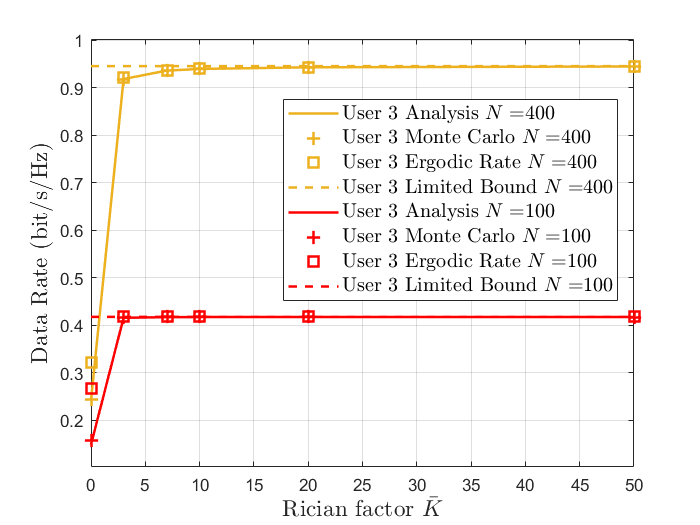}
    \caption{Monte Carlo Simulation vs. Derived closed-form expression with random weights, pilot allocation, \(K = 5\), \(\tau = 3\) , and \(|\mathcal{M}_k| =2\), \(\forall k\).}
    \label{Lower bound}
\end{figure}
In Fig. \ref{Lower bound}, the tightness between the derived lower bound and Monte-Carlo simulation is validated by averaging \(10^4\) trials. As expected, our derived expression perfectly matches the Monte-Carlo simulations. More importantly, the derived lower bound approaches the ergodic rate, especially for a large Rician factor. Furthermore, the data rate tends to be limited when \(\bar K_{m,k} \ge 20\), i.e., \(\bar K_{m,k} \ge 13\) dB, which confirms the analysis in (\ref{LosSINR}).

\subsection{Effectiveness of Algorithm {\ref{UGsolving}}}

\begin{figure}
    \centering
    \includegraphics[width=0.8\linewidth]{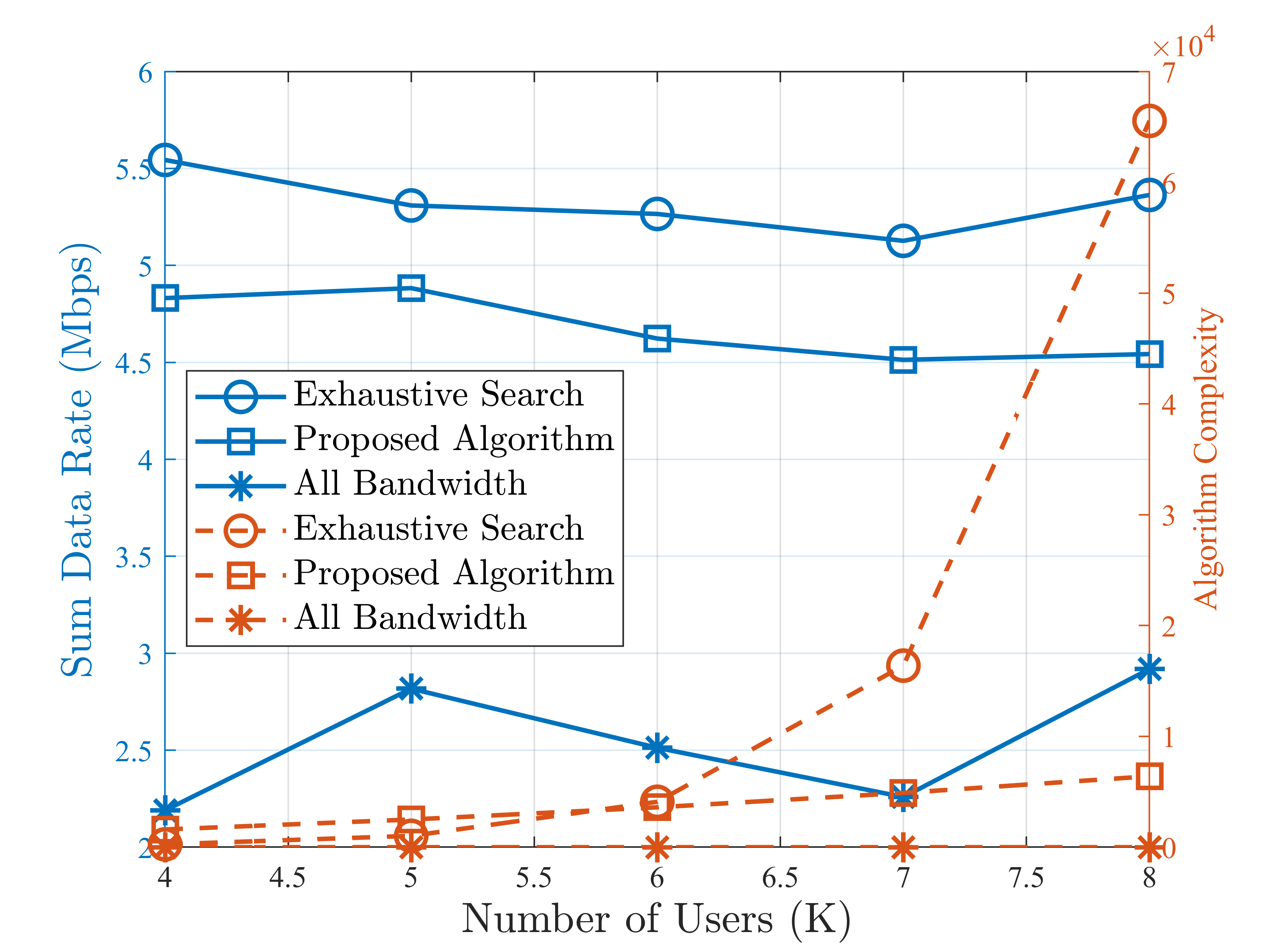}
    \caption{Algorithm \ref{UGsolving} vs. exhaustive search with random weights, pilot allocation, \(I = 4\), \(\tau = K -1 \) , and \(|\mathcal{M}_k| = 3\), \(\forall k\).}
    \label{Alg1}
\end{figure}

To validate the effectiveness of our proposed algorithm, we compare Algorithm {\ref{UGsolving}} with the exhaustive search, as shown in Fig. \ref{Alg1}. As can be seen, there is a gap between the proposed Algorithm {\ref{UGsolving}} and the exhaustive search, as our algorithm can only obtain a feasible solution rather than the optimal solution. In contrast, our algorithm achieves a significantly higher sum data rate compared to all bandwidth allocation scheme, where all users share the total bandwidth \(B\). Furthermore, as our expected, the proposed algorithm achieves lower computational complexity, especially for large numbers of users and sub-bands, thereby validating the efficacy of the user scheduling strategy in balancing performance and complexity.

\subsection{Convergence  of Proposed Algorithm}
\begin{figure}
    \centering
    \includegraphics[width=0.8\linewidth]{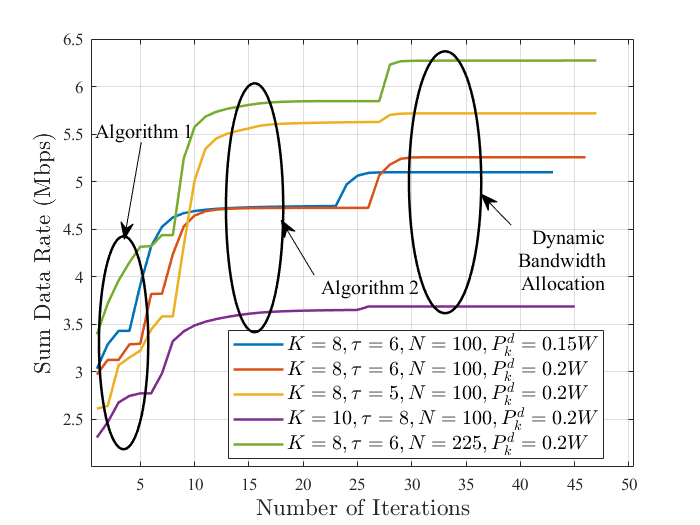}
    \caption{Convergence of our proposed method with random pilot allocation and \(|\mathcal{M}_k| = 3\) .}
    \label{convergence}
\end{figure}
Next, the convergence of our proposed method is illustrated in Fig. \ref{convergence}. By iteratively operating our proposed algorithms, we observe that our proposed method can converge to a locally optimal solution with only a few iterations (10 iterations for Algorithm 2 and 5 iterations for dynamic bandwidth allocation). Furthermore, increasing the number of antennas of each satellite can significantly improve the system performance. 

\subsection{Performace Comparison}
\begin{figure}
    \centering
    \includegraphics[width=0.8\linewidth]{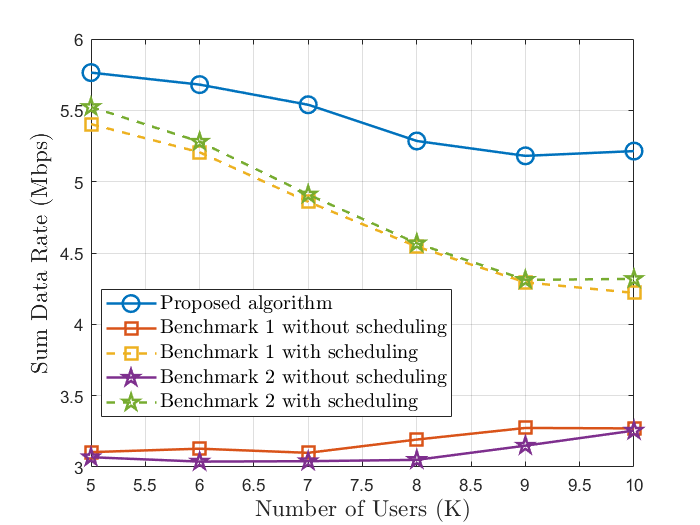}
    \caption{Performance comparison with random pilot allocation, \(I = 4\), \(\tau = K -2 \) , \(P^{d,\max}_k = 0.2 \) W, and \(|\mathcal{M}_k| = 3\), \(\forall k\).}
    \label{comparison}
\end{figure}

To demonstrate the effectiveness of our proposed method, the following benchmarks are considered:
\begin{enumerate}
    \item {\bf Benchmark 1}: We optimize the transmission power via Algorithm 2 with the equal weights \(w_{m,k} = \frac{1}{\sqrt{|\mathcal{M}_k|}}\), where \(|\mathcal{M}_k|\) is the cardinality of set \(\mathcal{M}_k\);
    \item {\bf Benchmark 2}: The weights are generated based on estimated channel, i.e., \(w_{m,k} = \frac{|\mathbf{\hat h}_{m,k}|}{\sqrt{\sum\limits_{m \in \mathcal{M}_k}|\mathbf{\hat h}_{m,k}|^2}}\), and the transmission power is optimized by using Algorithm 2;
\end{enumerate}
In Fig. \ref{comparison}, the data rate is averaged over 100 trials. As expected, our proposed method is superior to the two benchmarks, as it can judiciously adjust the user scheduling strategy, combining weights, transmission power, and bandwidth allocation, thereby improving the data rate. More importantly, by adopting the user strategy, the sum data rate of the distributed system can be significantly improved. This is due to the fact that the users with severe interference will not be allowed to share the common sub-band, thereby enhancing the data rate. Furthermore, our simulation results demonstrate a monotonic degradation in system performance as the number of users increases, even with advanced scheduling strategies. This trend arises from the close proximity of users, which induces highly correlated channels and severe inter-user interference.  Unlike scenarios with sparse deployments, the monotonic decrease underscores the fundamental limitation of spatial scheduling in dense environments, as interference saturates the system’s ability to distinguish users through spatial separation alone. More importantly, these simulation results reveal that this necessitates the adoption of time-division multiplexing or the allocation of extra orthogonal sub-bands to decouple interference-dominated users, thereby sustaining quality of service in densely populated satellite networks.

\section{Conclusion and Future Work}
In this paper, we first exploited the cooperative reception for the distributed MIMO systems while considering the dominant LoS channel and severe link budget. Then, by deriving the closed-form of estimation errors and the lower bound of the achievable data rate, the comprehensive performance analysis was conducted, revealing the impacts of channel estimation errors and pilot contamination on data rate. Furthermore, we observed that the severe inter-user interference significantly degrade the system performance. Thereafter, we maximized of sum data rate by dynamic resource allocation and then decomposed this NP-hard problem (i.e.,) into three tractable sub-problems, including user scheduling strategy, joint weights and power control, and dynamic bandwidth allocation. For the user scheduling strategy, we transformed it into a coloring graph and proposed a low-complexity algorithm based on the DSatura algorithm. After that, the joint weights and power control can be optimized by using SCA and GP. Finally, the optimal bandwidth allocation can be obtained by solving the concave problem. Simulation results validated our derivations and confirmed the effectiveness of our proposed method.

Considering the highly dense case, it is challenging to allocate the resources with the fixed number of sub-bands. Therefore, it is beneficial to explore how to split the bandwidth according to the users' requirements and the number of users. Owing to the resource-assignment problem, the conventional approach may no longer be suitable for the dynamical case, and thus the deep learning-based algorithm should be developed in the future.

\begin{appendices}
\section{Proof of Theorem \ref{MRC_SINR_T}}
\label{Prooftheorem1}

From (\ref{kth_SINR}), we need to derive the expressions of ${\left| {{\rm{DS}}_{k}} \right|^2}$, $\mathbb{E}\left( \left| {{\rm{LS}}_{k}} \right|^2 \right)$, $\mathbb{E}\left( \left| {{\rm{UI}}_{k,k'}} \right|^2 \right)$ and $\mathbb{E}\left( \left| {{\rm{N}}_{k}} \right|^2 \right)$, respectively.

We first rewrite the expression of the estimated channel \(\mathbf{\hat h}_{m,k}\) as
\begin{equation}
\small
    \label{rwestimatedchennel}
    \begin{split}
    &\mathbf{\hat h}_{m,k}\\
    =& \sqrt{\frac{{\bar K}_{m,k}}{{\bar K}_{m,k}+1}}\sqrt{\beta_{m,k}}\mathbf{\bar{h} }_{m,k}+\\
     & \sqrt{\tau p_k^p}\mathbf{R}_{m,k}\boldsymbol{\Psi}_{m,k} \Big(\sum_{j \in \mathcal{P}_k}\sqrt{\frac{\tau p^p_j\beta_{m,j}  }{{\bar K}_{m,j}+1}}\mathbf{\widetilde h}_{m,j} +{\bf{n}}_{m,k}^p \Big) \\
     = &{{\sqrt{\bar K_{m,k}a_{m,k}}\mathbf{\bar{h} }_{m,k}}}+  \\
    & \underbrace{\sqrt{\tau p_k^p}\mathbf{R}_{m,k}\boldsymbol{\Psi}_{m,k} \Big(\sum_{j \in \mathcal{P}_k}\sqrt{\tau p^p_j a_{m,j}}\boldsymbol{\Delta}_{m,j}^{\frac{1}{2}}\mathbf{\widetilde h}_{m,j} +{\bf{n}}_{m,k}^p \Big)}_{\mathbf{\widetilde q}_{m,k}}.
    \end{split}
\end{equation}
The channel \(\mathbf{ h}_{m,k}\) can be rewritten as
\begin{equation}
\small
    \mathbf{ h}_{m,k} = \sqrt{\bar K_{m,k}a_{m,k}}\mathbf{\bar{h} }_{m,k} + \sqrt{a_{m,k}}\boldsymbol{\Delta}_{m,k}^{\frac{1}{2}} \mathbf{\widetilde h}_{m,k}.
\end{equation}

Then, we compute ${\rm{DS}}_k$. Since estimated channel $\hat {\bf{h}}_{m,k}$ and estimation error $\mathbf{h}_{m,k}-\hat {\bf{h}}_{m,k}$ are independent, we have
\begin{equation}
\small
\label{MRC_DSk}
\begin{split}
	{{\rm{DS}}_{k}}  &= { \mathbb{E}{\left\{ {\sum\limits_{m \in {\mathcal{M}}_k} {\sqrt {{w_{m,k}p_k^d}} {{ {{\bf{\hat h}}^H_{m,k}} }} {{\bf{\hat h}}_{m,k}}} } \right\}} }  \\
	 &= {\sqrt{p_k^d}}  \sum\limits_{m \in {\mathcal{M}}_k} \sqrt{w_{m,k}}\Big( \tau p^p_k \text{tr}\{\mathbf{R}_{m,k}\boldsymbol{\Psi}_{m,k}\mathbf{R}_{m,k}\} \\
     &+ {\bar K}_{m,k}a_{m,k}||\mathbf{\bar{h} }_{m,k}||^2\Big) \\
\end{split}
\end{equation}

 Here, we note that \(\mathbf{\hat h}_{m,k}\) and \(\mathbf{\hat h}_{m,k'}\) are correlated if \(\{k,k'\} \in \mathcal{P}_k\). Therefore, we separately derive the interference's expression. First, if the \(k'\)-th user does not share the same pilot with the user \(k\), i.e., \(\{k,k'\} \notin \mathcal{P}_k\), then \(\mathbf{\hat h}_{m,k}\) and \(\mathbf{h}_{m,k'}\) are independent and we have
 \begin{equation}
 \small
\label{MRC_UIkk1}
\begin{split}
 &\mathbb{E} \left( {{{\left| {{\rm{U}}{{\rm{I}}_{k,k'}}} \right|}^2}} \right)\\
 =  &\mathbb{E} \left\{ {{{\left| {\sum\limits_{m \in {\mathcal{M}}_k} {\sqrt {w_{m,k}p_{k'}^d} {{\left( {{{{\bf{\hat h}}}_{m,k}}} \right)}^H}{{\bf{h}}_{m,k'}}} } \right|}^2}} \right\} \\
   =&  p^d_{k'} \left|\sum\limits_{m \in {\mathcal{M}}_k} \sqrt{w_{m,k}\bar K_{m,k}\bar K_{m,k'}a_{m,k}a_{m,k'}}\mathbf{\bar h}^H_{m,k}\mathbf{\bar h}^H_{m,k'}\right|^2 \\
   +& p^d_{k'} \mathbb{E}\left\{ \left| \sum\limits_{m \in \mathcal{M}_k} \sqrt{w_{m,k}\bar K_{m,k'}a_{m,k'}}\mathbf{\widetilde q}^H_{m,k}\mathbf{\bar h}_{m,k'}\right|^2\right\} \\
  + &  p^d_{k'} \mathbb{E}\left\{ \left|\sum\limits_{m \in {\mathcal{M}}_k}  \sqrt{w_{m,k}{\bar K}_{m,k}a_{m,k} a_{m,k'}} \mathbf{\bar h}^H_{m,k}\boldsymbol{\Delta}_{m,k'} \mathbf{\widetilde h}_{m,k'} \right|^2\right\} \\
  +& p^d_{k'} \mathbb{E}\left\{ \left| \sum\limits_{m \in {\mathcal{M}}_k} \sqrt{w_{m,k}a_{m,k'}}\mathbf{\widetilde q}^H_{m,k} \boldsymbol{\Delta}_{m,k'}^{\frac{1}{2}}\mathbf{\widetilde h}_{m,k'} \right|^2\right\}.
 \end{split}
 \end{equation}

The fist term in (\ref{MRC_UIkk1}) can be expressed as
\begin{equation}
\small
    \begin{split}
        &\mathbb{E}\left\{ \left| \sum\limits_{m \in \mathcal{M}_k} \sqrt{w_{m,k}\bar K_{m,k'}a_{m,k'}}\mathbf{\widetilde q}_{m,k}\mathbf{\bar h}_{m,k'}\right|^2\right\}\\
        = &\sum\limits_{m \in \mathcal{M}_k}w_{m,k} \bar K_{m,k'}a_{m,k'}\mathbb{E}\{\mathbf{\widetilde q}^H_{m,k}\mathbf{\bar h}_{m,k'}\mathbf{\bar h}^H_{m,k'} \mathbf{\widetilde q}_{m,k}\} \\
        = & \sum\limits_{m \in \mathcal{M}_k} w_{m,k}\bar K_{m,k'}\tau p^p_k a_{m,k'} \mathbf{\bar h}^H_{m,k'} \mathbf{R}_{m,k}\boldsymbol{\Psi}_{m,k} \mathbf{R}_{m,k}\mathbf{h}_{m,k'}.
    \end{split}
\end{equation}
The second term in (\ref{MRC_UIkk1}) can be given by
\begin{equation}
\small
    \begin{split}
        &\mathbb{E}\left\{ \left|\sum\limits_{m \in {\mathcal{M}}_k}  \sqrt{w_{m,k}{\bar K}_{m,k}a_{m,k} a_{m,k'}} \mathbf{\bar h}^H_{m,k} \boldsymbol{\Delta}^{\frac{1}{2}}_{m,k'} \mathbf{\widetilde h}_{m,k'} \right|^2\right\}\\
        = &\sum\limits_{m \in \mathcal{M}_k} w_{m,k}\bar K_{m,k}a_{m,k}a_{m,k'}\\
        \times &\mathbb{E}\{\mathbf{\bar h}^H_{m,k}\boldsymbol{\Delta}^{\frac{1}{2}}_{m,k'} \mathbf{\widetilde h}_{m,k'}\mathbf{\widetilde h}^H_{m,k'}\boldsymbol{\Delta}^{\frac{1}{2}}_{m,k'}  \mathbf{\bar h}_{m,k}\} \\
        = &\sum\limits_{m \in \mathcal{M}_k} w_{m,k}\bar K_{m,k}a_{m,k}a_{m,k'}\mathbf{\bar h}^H_{m,k}\boldsymbol{\Delta}_{m,k'} \mathbf{\bar h}_{m,k}.
    \end{split}
\end{equation}

The last term in (\ref{MRC_UIkk1}) is
\begin{equation}
\small
    \begin{split}
         &\mathbb{E}\left\{ \left| \sum\limits_{m \in {\mathcal{M}}_k} \sqrt{w_{m,k}a_{m,k'}}\mathbf{\widetilde q}^H_{m,k}\boldsymbol{\Delta}_{m,k'}^{\frac{1}{2}} \mathbf{\widetilde h}_{m,k'} \right|^2\right\} \\
       = & \sum\limits_{m \in \mathcal{M}_k} w_{m,k}a_{m,k'}\\
       \times &  \mathbb{E}\{\mathbf{\widetilde q}^H_{m,k} \boldsymbol{\Delta}_{m,k'}^{\frac{1}{2}}\mathbf{\widetilde h}_{m,k'}\mathbf{\widetilde h}^H_{m,k'}\boldsymbol{\Delta}_{m,k'}^{\frac{1}{2}}\mathbf{\widetilde q}_{m,k} \} \\
        = & \sum\limits_{m \in \mathcal{M}_k} w_{m,k}\tau p^p_k a_{m,k'}\text{tr}\{\boldsymbol{\Delta}_{m,k'}\mathbf{R}_{m,k}\boldsymbol{\Psi}_{m,k} \mathbf{R}_{m,k}\}.
    \end{split}
\end{equation}

 If  the \(k'\)-th user share the same pilot and sub-bandwidth with the user \(k\), i.e., \(\{k,k'\} \in \mathcal{P}_k\) and \(\{k,k'\} \in \mathcal{K}_i\), then \(\mathbf{\hat h}_{m,k}\) and \(\mathbf{h}_{m,k'}\) are correlated, and we have
\begin{equation}
\small
\label{MRC_UIkk2}
\begin{split}
 &\mathbb{E} \left( {{{\left| {{\rm{U}}{{\rm{I}}_{k,k'}}} \right|}^2}} \right)\\
 =  &\mathbb{E} \left\{ {{{\left| {\sum\limits_{m \in {\mathcal{M}}_k} {\sqrt {w_{m,k}p_{k'}^d} {{\left( {{{{\bf{\hat h}}}_{m,k}}} \right)}^H}{{\bf{h}}_{m,k'}}} } \right|}^2}} \right\} \\
   =&  p^d_{k'} \left|\sum\limits_{m \in {\mathcal{M}}_k} \sqrt{w_{m,k}\bar K_{m,k}\bar K_{m,k'}a_{m,k}a_{m,k'}}\mathbf{\bar h}^H_{m,k}\mathbf{\bar h}^H_{m,k'}\right|^2 \\
   +& p^d_{k'} \mathbb{E}\left\{ \left| \sum\limits_{m \in \mathcal{M}_k} \sqrt{w_{m,k}\bar K_{m,k'}a_{m,k'}}\mathbf{\widetilde q}^H_{m,k}\mathbf{\bar h}_{m,k'}\right|^2\right\} \\
  + &  p^d_{k'} \mathbb{E}\left\{ \left|\sum\limits_{m \in {\mathcal{M}}_k}  \sqrt{w_{m,k}{\bar K}_{m,k}a_{m,k} a_{m,k'}} \mathbf{\bar h}^H_{m,k}  \boldsymbol{\Delta}^{\frac{1}{2}}_{m,k'}\mathbf{\widetilde h}_{m,k'} \right|^2\right\} \\
  +& p^d_{k'} \mathbb{E}\left\{ \left| \sum\limits_{m \in {\mathcal{M}}_k} \sqrt{w_{m,k}a_{m,k'}}\mathbf{\widetilde q}^H_{m,k}  \boldsymbol{\Delta}^{\frac{1}{2}}_{m,k'}\mathbf{\widetilde h}_{m,k'} \right|^2\right\}\\
  +& 2 p^d_{k'} \text{Re}\Big\{ \sum\limits_{m \in \mathcal{M}_k}\sqrt{w_{m,k}\bar K_{m,k}a_{m,k}\bar K_{m,k'}a_{m,k'}}\mathbf{\bar h}^H_{m,k}\mathbf{\bar{h} }_{m,k'} \\
  \times & \sum\limits_{n \in \mathcal{M}_k} \sqrt{w_{n,k} a_{n,k'}}\mathbb{E}\{\mathbf{\widetilde h}^H_{n,k'} \boldsymbol{\Delta}^{\frac{1}{2}}_{n,k'}\mathbf{\widetilde q}_{n,k}\}\Big \}.
 \end{split}
 \end{equation}
In the following, we calculate the terms in (\ref{MRC_UIkk2}). The second and third terms are derived, and thus we derive the term \( \mathbb{E}\left\{ \left| \sum\limits_{m \in {\mathcal{M}}_k} \sqrt{w_{m,k}a_{m,k'}}\mathbf{\widetilde q}^H_{m,k}  \boldsymbol{\Delta}^{\frac{1}{2}}_{m,k'}\mathbf{\widetilde h}_{m,k'} \right|^2\right\}\), which is given by
\begin{equation}
\small
\label{UIKK2c}
    \begin{split}
    &\mathbb{E}\left\{ \left| \sum\limits_{m \in {\mathcal{M}}_k} \sqrt{w_{m,k}a_{m,k'}}\mathbf{\widetilde q}^H_{m,k}  \boldsymbol{\Delta}^{\frac{1}{2}}_{m,k'}\mathbf{\widetilde h}_{m,k'} \right|^2\right\}\\
    =& \sum\limits_{m \in {\mathcal{M}}_k}\sum\limits_{n \in {\mathcal{M}}_k} \sqrt{w_{m,k}w_{n,k}a_{m,k'}a_{n,k'}}\\
    \times &\mathbb{E}\{\mathbf{\widetilde q}^H_{m,k}  \boldsymbol{\Delta}^{\frac{1}{2}}_{m,k'}\mathbf{\widetilde h}_{m,k'}\mathbf{\widetilde h}^H_{n,k'}  \boldsymbol{\Delta}^{\frac{1}{2}}_{n,k'}\mathbf{\widetilde q}_{n,k}\} \\
    =& \sum\limits_{m \in {\mathcal{M}}_k}\sum\limits_{n \neq m} \sqrt{w_{m,k}w_{n,k}a_{m,k'}a_{n,k'}} \\
    \times & \mathbb{E}\{\mathbf{\widetilde q}^H_{m,k}  \boldsymbol{\Delta}^{\frac{1}{2}}_{m,k'}\mathbf{\widetilde h}_{m,k'}\mathbf{\widetilde h}^H_{n,k'} \boldsymbol{\Delta}^{\frac{1}{2}}_{n,k'} \mathbf{\widetilde q}_{n,k} \}\\
    +  &\sum\limits_{m \in {\mathcal{M}}_k} w_{m,k}a_{m,k'}\times\\
    &\mathbb{E}\{\mathbf{\widetilde q}^H_{m,k} \boldsymbol{\Delta}^{\frac{1}{2}}_{m,k'} \mathbf{\widetilde h}_{m,k'}\mathbf{\widetilde h}^H_{m,k'}  \boldsymbol{\Delta}^{\frac{1}{2}}_{m,k'}\mathbf{\widetilde q}_{m,k}\}
    \end{split}
\end{equation}

The first term in (\ref{UIKK2c}) is given by 
\begin{equation}
\small
    \begin{split}
       & \mathbb{E}\{\mathbf{\widetilde q}^H_{m,k}  \boldsymbol{\Delta}^{\frac{1}{2}}_{m,k'}\mathbf{\widetilde h}_{m,k'}\mathbf{\widetilde h}^H_{n,k'} \boldsymbol{\Delta}^{\frac{1}{2}}_{n,k'} \mathbf{\widetilde q}_{n,k} \} \\
        =& \tau^2 p^p_k p^p_{k'}\sqrt{a_{m,k'}a_{n,k'}} \text{tr}\{ \boldsymbol{\Delta}_{m,k'} \mathbf{R}_{m,k}\boldsymbol{\Psi}_{m,k} \} \\
        \times &\text{tr}\{ \boldsymbol{\Delta}_{n,k'} \mathbf{R}_{n,k}\boldsymbol{\Psi}_{n,k} \} .
    \end{split}
\end{equation}

The second term in (\ref{UIKK2c}) can be calculated as
\begin{equation}
\small
    \begin{split}
    & \mathbb{E}\{\mathbf{\widetilde q}^H_{m,k} \boldsymbol{\Delta}^{\frac{1}{2}}_{m,k'} \mathbf{\widetilde h}_{m,k'}\mathbf{\widetilde h}^H_{m,k'}  \boldsymbol{\Delta}^{\frac{1}{2}}_{m,k'}\mathbf{\widetilde q}_{m,k}\}\\
    =&  \tau p_k^p \mathbb{E}\Big\{\sum_{j \in \mathcal{P}_k} \tau p^p_ja_{m,j} \mathbf{\widetilde h}^H_{m,j}\boldsymbol{\Delta}^{\frac{1}{2}}_{m,j} \boldsymbol{\Psi}_{m,k}\mathbf{R}_{m,k} \boldsymbol{\Delta}^{\frac{1}{2}}_{m,k'}\\
    \times& \mathbf{\widetilde h}_{m,k'}\mathbf{\widetilde h}^H_{m,k'}\boldsymbol{\Delta}^{\frac{1}{2}}_{m,k'}\mathbf{R}_{m,k}\boldsymbol{\Psi}_{m,k}\boldsymbol{\Delta}^{\frac{1}{2}}_{m,j}\mathbf{\widetilde h}_{m,j} \Big\} \\
    + &\tau p_k^p |{(\bf{n}}_{m,k}^p)^H\boldsymbol{\Psi}_{m,k}\mathbf{R}_{m,k} \boldsymbol{\Delta}^{\frac{1}{2}}_{m,k'} \mathbf{ \widetilde h}_{m,k'}|^2 \\
    = & \tau p_k^p \sigma^2 \text{tr}\{\boldsymbol{\Psi}_{m,k}\mathbf{R}_{m,k} \boldsymbol{\Delta}_{m,k'}\mathbf{R}_{m,k}\boldsymbol{\Psi}_{m,k}\} \\
    +& \tau p_k^p \sum\limits_{j \in \mathcal{P}_k}\tau p^p_j a_{m,j}\text{tr}\{ \boldsymbol{\Delta}_{m,j}\boldsymbol{\Psi}_{m,k}\mathbf{R}_{m,k} \boldsymbol{\Delta}_{m,k'}\mathbf{R}_{m,k}\boldsymbol{\Psi}_{m,k}\} \\
    +& \tau^2 p_k^p p_{k'}^pa_{m,k'}\text{tr}\{\boldsymbol{\Delta}_{m,k'}\boldsymbol{\Psi}_{m,k}\mathbf{R}_{m,k} \}\text{tr}\{\boldsymbol{\Delta}_{m,k'}\boldsymbol{\Psi}_{m,k}\mathbf{R}_{m,k} \} \\
    = &\tau^2 p_k^p p_{k'}^pa_{m,k'} \text{tr}\{\boldsymbol{\Delta}_{m,k'}\boldsymbol{\Psi}_{m,k}\mathbf{R}_{m,k} \}\text{tr}\{\boldsymbol{\Delta}_{m,k'}\boldsymbol{\Psi}_{m,k}\mathbf{R}_{m,k} \}\\
    +& \tau p_k^p \text{tr}\{\mathbf{R}_{m,k} \boldsymbol{\Delta}_{m,k'}\mathbf{R}_{m,k}\boldsymbol{\Psi}_{m,k}\} .   
    \end{split}
\end{equation}

\(\mathbb{E}\{\mathbf{\bar h}^H_{m,k}\mathbf{\bar{h} }_{m,k'} \mathbf{\widetilde h}^H_{n,k'} \boldsymbol{\Delta}^{\frac{1}{2}}_{n,k'}\mathbf{\widetilde q}_{n,k}\}\) can be calculated as
\begin{equation}
\small
    \begin{split}
        &\mathbb{E}\{\mathbf{\bar h}^H_{m,k}\mathbf{\bar{h} }_{m,k'} \mathbf{\widetilde h}^H_{n,k'} \boldsymbol{\Delta}^{\frac{1}{2}}_{n,k'}\mathbf{\widetilde q}_{n,k}\} \\
        = & \tau\sqrt{ p^p_{k'}p^p_{k}a_{n,k'}} \mathbf{\bar h}^H_{m,k}\mathbf{\bar{h} }_{m,k'} \text{tr}\{\boldsymbol{\Delta}_{n,k'}\mathbf{R}_{n,k}\boldsymbol{\Psi}_{n,k}\}.
    \end{split}
\end{equation}

The term $\mathbb{E}\left( \left| {{\rm{LS}}_{k}} \right|^2 \right)$ is given by
\begin{equation}
\setlength\abovedisplayskip{5pt}
\setlength\belowdisplayskip{5pt}
\label{MRC_LSk}
\small
\begin{split}
&\mathbb{E} \left\{ {{{\left| {{\rm{LS}}_{k}} \right|}^2}} \right\}  
= {p_k^d} \mathbb{E} \left\{ {{{\left| {\sum\limits_{m \in {\mathcal{M}}_k}{{{\left( {{\bf{\hat h}}_{m,k}} \right)}^H}{\bf{h}}_{m,k} } -  {{\rm{DS}}_{k}}} \right|}^2}} \right\}   \\
= &{p_k^d}\Big(\sum\limits_{m \in {\mathcal{M}}_k} \sqrt{w_{m,k}\bar K_{m,k}a_{m,k}}\mathbf{\widetilde q}^H_{m,k} \mathbf{\bar{h} }_{m,k} \\
+& \sum\limits_{m \in {\mathcal{M}}_k}\sqrt{w_{m,k}\bar K_{m,k}a_{m,k}a_{m,k}} \mathbf{\bar{h} }^H_{m,k}\boldsymbol{\Delta}_{m,k}^{\frac{1}{2}}\mathbf{\widetilde{h} }_{m,k}\\
  +&\sum\limits_{m \in {\mathcal{M}}_k}\sqrt{w_{m,k}a_{m,k}}\mathbf{\widetilde q}^H_{m,k}\boldsymbol{\Delta}_{m,k}^{\frac{1}{2}}\mathbf{\widetilde{h} }_{m,k} \\
  -& \sum\limits_{m \in {\mathcal{M}}_k}\sqrt{w_{m,k}}\tau p^p_k \text{tr}\{\mathbf{R}_{m,k}\boldsymbol{\Psi}_{m,k}\mathbf{R}_{m,k}\}\Big)^2 \\
 =& {p_k^d} \sum\limits_{m \in {\mathcal{M}}_k} w_{m,k}\bar K_{m,k}a_{m,k} \mathbb{E}\{|\mathbf{\widetilde q}^H_{m,k}\mathbf{\bar{h} }_{m,k} |^2\} \\
 +& {p_k^d} \sum\limits_{m \in {\mathcal{M}}_k}  w_{m,k}\bar K_{m,k}a_{m,k}a_{m,k} \mathbb{E}\{|\mathbf{\bar{h} }^H_{m,k}\boldsymbol{\Delta}_{m,k}^{\frac{1}{2}}\mathbf{\widetilde{h} }_{m,k} |^2\}\\
  + & {p_k^d} \mathbb{E}\Big\{ \sum\limits_{m \in {\mathcal{M}}_k}\sqrt{w_{m,k}a_{m,k} }\mathbf{\widetilde q}^H_{m,k}\boldsymbol{\Delta}_{m,k}^{\frac{1}{2}}\mathbf{\widetilde{h} }_{m,k} \\ 
  \times& \sum\limits_{n \in {\mathcal{M}}_k} \sqrt{w_{n,k} a_{n,k}}\mathbf{\widetilde{h} }^H_{n,k}\boldsymbol{\Delta}_{n,k}^{\frac{1}{2}}\mathbf{\widetilde q}_{n,k}\Big\}\\
  + &{p_k^d}\Big(\sum\limits_{m \in {\mathcal{M}}_k}\sqrt{w_{m,k}}\tau p^p_k \text{tr}\{\mathbf{R}_{m,k}\boldsymbol{\Psi}_{m,k}\mathbf{R}_{m,k}\}\Big)^2 \\
  - & 2{p_k^d}\text{Re}\Big\{\sum\limits_{m \in {\mathcal{M}}_k}\sum\limits_{n \in {\mathcal{M}}_k}  \tau p^p_k \text{tr}\{\mathbf{R}_{n,k}\boldsymbol{\Psi}_{n,k}\mathbf{R}_{n,k}\}\\
  \times&\sqrt{w_{m,k}w_{n,k}a_{m,k}}\mathbb{E}\{\mathbf{\widetilde q}^H_{m,k}\boldsymbol{\Delta}_{m,k}^{\frac{1}{2}}\mathbf{\widetilde{h} }_{m,k}\}\Big\} .
 \end{split}
\end{equation}

Then, the first term \(\mathbb{E}\{|\mathbf{\widetilde q}^H_{m,k}\mathbf{\bar{h} }_{m,k} |^2\}\) of (\ref{MRC_LSk}) can be calculated as
\begin{equation}
\small
    \begin{split}
        \mathbb{E}\{|\mathbf{\widetilde q}^H_{m,k}\mathbf{\bar{h} }_{m,k} |^2\} = \tau p^p_k \mathbf{\bar h}^H_{m,k} \mathbf{R}_{m,k}\boldsymbol{\Psi}_{m,k}\mathbf{R}_{m,k}\mathbf{\bar h}_{m,k}.
    \end{split}
\end{equation}
The second term \(\mathbb{E}\{|\mathbf{\bar{h} }^H_{m,k}\mathbf{\widetilde{h} }_{m,k} |^2\}\) is given by
\begin{equation}
\small
    \mathbb{E}\{|\mathbf{\bar{h} }^H_{m,k}\mathbf{\widetilde{h} }_{m,k} |^2\} = \mathbf{\bar h}^H_{m,k}\boldsymbol{\Delta}_{m,k}\mathbf{\bar h}_{m,k}.
\end{equation}
The third term in (\ref{MRC_LSk}) can be written as
\begin{equation}
\small
\begin{split}
    &\mathbb{E}\Big\{ \sum\limits_{m \in {\mathcal{M}}_k}\sqrt{w_{m,k}a_{m,k} }\mathbf{\widetilde q}^H_{m,k}\boldsymbol{\Delta}_{m,k}^{\frac{1}{2}}\mathbf{\widetilde{h} }_{m,k} \\ &\times \sum\limits_{n \in {\mathcal{M}}_k} \sqrt{w_{n,k} a_{n,k}}\mathbf{\widetilde{h} }^H_{n,k}\boldsymbol{\Delta}_{n,k}^{\frac{1}{2}}\mathbf{\widetilde q}_{n,k}\Big\}  \\
    =&  \Big(\tau p^p_k \sum \limits_{m \in \mathcal{M}_k}w_{m,k} a_{m,k}\text{tr}\{\boldsymbol{\Delta}_{m,k}\mathbf{R}_{m,k}\boldsymbol{\Psi}_{m,k}   \} \Big)^2 \\
    + &\sum\limits_{m \in \mathcal{M}_k} w_{m,k}\tau p_k^p a_{m,k}\text{tr}\{\mathbf{R}_{m,k} \boldsymbol{\Delta}_{m,k}\mathbf{R}_{m,k}\boldsymbol{\Psi}_{m,k}\}. 
\end{split} 
\end{equation}
The fourth term \(\mathbb{E}\{\mathbf{\widetilde q}^H_{m,k}\boldsymbol{\Delta}_{m,k}^{\frac{1}{2}}\mathbf{\widetilde{h} }_{m,k}\}\) is calculated as
\begin{equation}
\small
   \mathbb{E}\{\mathbf{\widetilde q}^H_{m,k}\boldsymbol{\Delta}_{m,k}^{\frac{1}{2}}\mathbf{\widetilde{h} }_{m,k}\} = \tau p^p_k \sqrt{a_{m,k}} \text{tr}\{\boldsymbol{\Delta}_{m,k}\mathbf{R}_{m,k}\boldsymbol{\Psi}_{m,k}\}.
\end{equation}
By substituting all terms into (\ref{MRC_LSk}), we have
\begin{equation}
\small
    \begin{split}
       &  \mathbb{E} \left\{ {{{\left| {{\rm{LS}}_{k}} \right|}^2}} \right\}\\
        = &   {p_k^d} \sum\limits_{m \in \mathcal{M}_k}w_{m,k} \bar K_{m,k}\tau p^p_k a_{m,k} \mathbf{\bar h}^H_{m,k} \mathbf{R}_{m,k}\boldsymbol{\Psi}_{m,k} \mathbf{R}_{m,k}\mathbf{h}_{m,k}\\
        + & p_k^d \sum\limits_{m \in \mathcal{M}_k} w_{m,k}\bar K_{m,k}a_{m,k}a_{m,k}\mathbf{\bar h}^H_{m,k}\boldsymbol{\Delta}_{m,k} \mathbf{\bar h}_{m,k}\\
        + & p_k^d \sum\limits_{m \in \mathcal{M}_k} w_{m,k}\tau p_k^p a_{m,k}\text{tr}\{\mathbf{R}_{m,k} \boldsymbol{\Delta}_{m,k}\mathbf{R}_{m,k}\boldsymbol{\Psi}_{m,k}\}
    \end{split}
\end{equation}

Finally, we compute $\mathbb{E}\left( \left| {{\rm{N}}_{k}} \right|^2 \right)$, which is given by
\begin{equation}
\label{MRC_Nk}
\small
\begin{split}
    \mathbb{E} \left( {{{\left| {{{\rm{N}}_k}} \right|}^2}} \right) =& \mathbb{E}\{|\sum\limits_{m \in \mathcal{M}_k} \sqrt{w_{m,k}} {{{\left( {{\bf{ \hat h}}_{m,k}} \right)}^H}{\bf{n}}_m}|^2\} \\
         = &\sum\limits_{m \in \mathcal{M}_k}{w_{m,k}}\Big( \tau p^p_k  \text{tr}\{\mathbf{R}_{m,k}\boldsymbol{\Psi}_{m,k}\mathbf{R}_{m,k}\} \\
        + &\bar K_{m,k}a_{m,k} \mathbf{\bar h}^H_{m,k}\mathbf{\bar h}_{m,k}\Big)\sigma^2 .
\end{split}
\end{equation}

Substituting all these results into (\ref{kth_SINR}), we obtain \(1 \) in (\ref{MRC_SINR_LB}).
\end{appendices}


\bibliographystyle{IEEEtran}
\bibliography{myref}

\end{document}